\documentclass[twocolumn,showpacs,amsmath,amssymb,superscriptaddress]{revtex4-2}
\usepackage{graphicx}
\usepackage{hyperref}
\usepackage{xcolor}
\usepackage{color}
\usepackage[all,cmtip]{xy}
\usepackage{amsmath,amsthm} 
\usepackage{amsfonts}
\usepackage{multirow}
\usepackage{textcomp}

\makeatother

\begin{document}

\title{Axion insulators protected by $C_2\mathbb{T}$ and their K-theory invariants and material realization.}

\author{Rafael Gonz\'{a}lez-Hern\'{a}ndez}
\email{rhernandezj@uninorte.edu.co}
\affiliation{Departamento de F\'{i}sica y Geociencias, Universidad del Norte, Km. 5 V\'{i}a Antigua Puerto Colombia, Barranquilla 080020, Colombia}
\author{Carlos Pinilla}
\email{ccpinilla@uninorte.edu.co}
\affiliation{Departamento de F\'{i}sica y Geociencias, Universidad del Norte, Km. 5 V\'{i}a Antigua Puerto Colombia, Barranquilla 080020, Colombia}
\author{Bernardo Uribe}
\email{bjongbloed@uninorte.edu.co }
\affiliation{Departamento de Matem\'{a}ticas y Estad\'{i}stica, Universidad del Norte, Km. 5 V\'{i}a Antigua Puerto Colombia, Barranquilla 080020, Colombia}

\date{\today}

\begin{abstract}
Axion insulators are generally understood as magnetic topological insulators whose Chern-Simons axion coupling term is quantized and equal to $\pi$. Inversion and time reversal, or the composition of either one with a rotation or a translation, are symmetries which protect this invariant. In this work we focus our attention on the composition of a 2-fold rotation with time reversal, and we show that insulators with this symmetry possess a $\mathbb{Z}_2$ invariant arising from Atiyah's real K-theory. We call this invariant the K-theory Kane-Mele invariant due to the similarities it has with the Kane-Mele invariant for systems with time reversal symmetry. Whenever all Chern numbers vanish, we demonstrate that this invariant is equivalent to the Chern-Simons axion coupling, and in the presence of the inversion symmetry, 
we show how this invariant could be obtained from the eigenvalues of the inversion operator on its fixed points in momentum space.
For the general case of non-trivial Chern numbers,  the Chern-Simons axion coupling term incorporates
information of the K-theory Kane-Mele invariant as well as information regarding bands with non-trival Chern number.
An explicit formula in terms of K-theory generators is presented for the Chern-Simons axion coupling term, the
relation with the K-theory Kane-Mele invariant is explained, and a formula in terms of eigenvalues of the inversion
operator is obtained.
 Using an effective Hamiltonian model and first-principles calculations, we also show that the occurrence of bulk-band inversion and nontrivial K-theory Kane-Mele invariant index can be observed in axion insulators of the pnictides family. In particular, we demonstrate that NpBi can be classified as an axion insulator due to the detection of additional topological indicators such as the quantum spin Hall effect, gapped surface states, surface quantized anomalous Hall effect, and chiral hinge modes.

\end{abstract}
\maketitle
	
\section{Introduction}

The interplay between magnetism and electrical properties through magnetoelectric effects in quantum materials has recently been a topic of intense theoretical and experimental research.
Specially, the use of topology to describe new phases of matter has led to improve the understanding of many condensed matter systems \cite{RevModPhys.83.1057,RevModPhys.88.021004}.
This is the case of the magnetoelectric response of solids, which is expected to be quantized for exotic topological phases of matter \cite{Magnetoelectric_Polarizability_and_Axion_Electrodynamics_in_Crystalline_Insulators, Wieder}.   
In particular, topological insulators are distinguished from trivial insulators by their distinguished electromagnetic response, which can be described by the so-called Chern-Simons axion coupling term or $\theta$-term \cite{PhysRevB.92.085113}. The value of $\theta$ is defined only module 2$\pi$ and it is mapped independently to $-\theta$ for time reversal symmetry (TRS), inversion symmetry, or any composition of the previous two with rotations
{or translations} \cite{axion-physics, Axion_electrodynamics, Magnetic_topological_quantum_chemistry, Axion-coupling-in-the-hybrid-Wannier}. 
This property allows the $Z_2$ classification with $\theta$ values of $0$ or $\pi$ for trivial or strong-topological insulators (TI) respectively \cite{PhysRevB.78.195424}. In TIs the TRS symmetry protects the quantum spin Hall surface states, as it is the case for the material Bi$_2$Se$_3$ \cite{Bi2Se3}.

For materials that break TRS, but preserve inversion symmetry or TRS composed with rotations {or translations}, the $\theta$ term is expected to be quantized. 
In the case of magnetic materials with inversion symmetry, the quantization of $\theta$ indicates that the material
is an axion insulator \cite{axion-physics,  Axion_electrodynamics, Magnetic_topological_quantum_chemistry, Surfaces_of_axion_insulators}. 
In these materials TRS is also broken at the surfaces, and subsequently, the materials may behave like an insulator in these regions. 
This property can yield unique phenomena that are characteristic of axion insulators, such as the half-quantized surface anomalous Hall effect or the quantized topological magnetoelectric effect \cite{Surfaces_of_axion_insulators, Axion_electrodynamics}. 

Material realization of axion insulators has been scarce and elusive, although some
 examples of potential axion insulators have been proposed in recent years \cite{axi-sandwich}. 
In particular, magnetic heterostructures of Cr-doped (Bi,Sb)$_2$Te$_3$ thin films have been shown to behave as
 an axion insulator by observing a zero Hall plateau in the electronic transport measurements \cite{Mogi2017}. 
Theoretical and experimental measurements have demonstrated that MnBi$_2$Te$_4$, MnBi$_6$Te$_{10}$ and MnBi$_8$Te$_{13}$ materials could hold the axion insulator phase in zero magnetic fields \cite{septuple-layer-MnBi2Te4, MnBi2Te4-Gu, MnBi2Te4-family, MnBi2Te4-jing, MnBi6Te10, MnBi6Te10-tian, MnBi8Te13,MnBi8Te13-zhong}. 
Recently $f$-based antiferromagnetic candidates have been also suggested, namely EuIn$_2$As$_2$, EuSn$_2$As$_2$, and EuSn$_2$P$_2$ \cite{EuIn2As2-prr,EuIn2As2-prl,EuSn2As2,EuSn2As2-lv,EuSn2P2-Gui, EuSn2P2-pnas}. All of them are classified with the $Z_4$=2 topological invariant index.

On the other hand, mono pnictides of neptunium NpX (with X=As,Sb,Bi) are materials that crystallize at ambient conditions in the cubic-NaCl structure and that are known to display an antiferromagnetic ground state with a characteristic noncollinear 3{\bf k} structure of ordered magnetic moments parallel to the family of wave vector $\mathbf{k}=(111)$\cite{LANDER19957}. 
In this work we have identified NpBi and NpSb as potential materials to contain axion insulator properties as predicted in \cite{High-throughput}. 
To characterize this behavior in NpBi and NpSb we have found several axion insulator indicators such as band inversion, quantum spin hall effect, K-theory Kane-Mele index, surface quantized anomalous Hall effect and chiral hinge states. 
Given the similar effect on these two monopnictides, we will concentrate our discussion on the case of NpBi but the results extend directly to NpSb. 

We start our work with a characterization of the {K-theory invariants whenever a $C_2\mathbb{T}$ symmetry is present.}
Here we have focused our attention on the case in which a 2-fold rotation composed of TRS is preserved but
TRS is broken. We show that systems preserved by this symmetry possess a $\mathbb{Z}_2$ invariant coming from Atiyah's real K-theory; we have coined this invariant as the {\it K-theory Kane-Mele invariant}. We chose
this name because it is similar to the K-theory version of the Kane-Mele invariant on systems that possess TRS.
Then we demonstrate that the Chern-Simons axion coupling, which determines the behavior of the magnetoelectric tensor, is equivalent
to the K-theory Kane-Mele for systems with $C_2 \mathbb{T}$ symmetry {whenever all Chern numbers vanish}. Finally, we find that in the presence of the inversion symmetry, the 
Chern-Simons axion coupling term can be read from the eigenvalues of the inversion operator on its fixed points in momentum space. Whenever the Chern numbers do not vanish, the Chern-Simons axion coupling term incorporates
information of the K-theory Kane-Mele invariant, as well as information from bands with non-trivial Chern numbers coupled
 with bands with non-trivial winding numbers along paths. The explicit formula for the Chern-Simons axion coupling term based on
  the complete K-theory invariants is presented, the relation between the K-theory Kane-Mele invariant is explained, and
   a formula in terms of the negative eigenvalues of the inversion operator is deduced.
   We finish this first part with a summary of our findings emphasizing the relation between the K-theory Kane-Mele invariant
and the Chern-Simons axion coupling term.

Having understood the topological invariants of axion insulators, we proceed to present our candidate NpBi
for the axion insulators. We show that indeed it possesses the non-trivial  K-theory Kane-Mele invariant 
and that this invariant is equivalent to the Chern-simons axion coupling term,
and more importantly, that has both half-quantized anomalous Hall conductance and chiral hinge states.


\section{Topology of axion insulators}

Axion insulators are materials where time reversal symmetry $\mathbb{T}$ is broken
and where the Chern-Simons axion coupling 
\begin{align} \label{Chern-Simons axion coupling}
\theta = \frac{1}{4\pi} \int_{\mathrm{BZ}} d^3k \epsilon^{\alpha \beta \gamma} 
\mathrm{Tr} \left( \mathcal{A}_\alpha \partial_\beta \mathcal{A}_{\gamma} - i \frac{2}{3} \mathcal{A}_\alpha \mathcal{A}_\beta \mathcal{A}_\gamma \right)
\end{align}
is non-trivial with $\theta = \pi$ and its protected by another symmetry \cite{Surfaces_of_axion_insulators}. 
This quantity is only defined modulo $2 \pi$ and plays the role of the 3D holonomy whenever {the integral of} the second Chern class is trivial, 
as it was shown originally by \citet{Chern-Simons}. Its relevance comes from the fact that it describes an isotropic 
contribution of the magnetoelectric tensor $\alpha_{ij}= (\partial P_i / \partial B_j)_E$ \cite{Magnetoelectric_Polarizability_and_Axion_Electrodynamics_in_Crystalline_Insulators}, under which
an electric field will induce a parallel magnetization with a proportionality constant given by 
\begin{align}
\alpha_{\mathrm{iso}} = \frac{e^2}{\hbar} \frac{\theta}{2\pi}.
\end{align}
 
 The quantity $\theta$ is quantized to be $0$ or $\pi$ whenever time reversal, inversion
 or any composition of a rotation or translation with either time reversal or inversion is preserved. This follows
 from the facts that time reversal transforms $B_j$ into $-B_j$ and inversion $P_i$ into $-P_i$, and therefore
 each one of them transforms $\theta$ into $-\theta$. Rotations and translations, on the other hand, leave $\theta$ fixed, but
  composed with either time reversal or inversion they flip the sign of $\theta$. Hence the symmetries that can quantize
  $\theta$ are $ YX$ where $Y= \mathbb{T}, I$ and $X$ is any rotation $X=C_2, C_3,C_4,C_6$
or any translation.
  Note that whenever a system preserves $YC_3$, then $(YC_3)^3 = \pm Y $, and therefore
  $Y$ is also preserved. Similarly, whenever $YC_6$ is preserved, then $(YC_6)^3 = \pm YC_2$,
  and therefore $YC_2$ is also preserved. Hence the symmetries {not involving translations} that quantize $\theta$ can be
  taken to be: $\mathbb{T}$, $I$, $C_2\mathbb{T}$, $C_2I$, $C_4\mathbb{T}$ and $C_4I$.
  
 In what follows we will study the K-theory invariants of the symmetries 
 $\mathbb{T}$ and  $C_2\mathbb{T}$, and we will show that the 
 K-theory invariant associated to the bulk is equivalent to the Chern-Simons axion coupling.
 We just need only to note that there are two types of $C_2$ rotations on crystals, either it is a rotation
 around a coordinate axis $C_{2w}$ with $w=x,y,z$, or a rotation around  diagonal $C_{2{\bf d}}$ with
 ${\bf d} \in \{ ( 1\bar{1}0 ), ( 01\bar{1} ), ( \bar{1}01 ) \} $.
 The case $C_{2w}\mathbb{T}$ has been studied before \cite{Symmetry_representation_approach_to_topological_invariants}, but the case $C_{2{\bf d}} \mathbb{T}$
 needs particular consideration. Their properties are similar but their complete topological invariants are different.

\subsection{KR-theory invariants}

Let $\mathbb{R}^{a,b} := \mathbb{R}^a \oplus  \mathbb{R}^b$ endowed with the involution
\begin{align}
\tau(x_1,\dots,x_a,x_{a+1},\dots,x_{a+b}) &= \\ (x_1,\dots,x_a,& -x_{a+1},\dots,-x_{a+b}). \nonumber
\end{align}
 Denote $B^{a,b}$ the ball of vectors
of norm less or equal than 1 and denote $S^{a,b}:=B^{a,b}/\partial B^{a,b}$ the one point
compactification of the boundary. The spheres $S^{a,b}$ are then topological spaces with involution.

Let us now recall the definition of real K-theory $KR$ of \citet{Atiyah}. For a space $X$ with involution $\tau$, a real 
vector bundle $E \to X$ consists of a complex vector bundle $E$ endowed with an involution $J$ lifting $\tau$
such that $J$ acts antiunitarily on the fibers ($J:E_x \to E_{\tau(x)}$ is complex anti-linear) and such that $J^2=1$.
Let $KR(X)$ be the Grothendieck group of real vector bundles over $X$. The relative group $KR(X,Y)$ is defined 
as $\widetilde{KR}(X/Y)$ where the relative $KR$ group $\widetilde{KR}$ is defined as the kernel to the restriction to base point.

The suspension groups are
\begin{align}
KR^{a,b}(X,Y) = KR(X \times B^{a,b}, X \times \partial B^{a,b} \cup Y \times B^{a,b}),
\end{align}
and therefore the usual suspension groups permit us to define the negative $KR$-groups as
\begin{align}
KR^{-a}=KR^{a,0}.
\end{align}
Multiplication by the Bott class 
\begin{align} \label{Bott class}
[H]-1 \in KR^{1,1}(*) = \widetilde{KR}(S^{1,1})
\end{align}
induces the Bott isomorphism \cite[Thm. 2.3]{Atiyah}
\begin{align}
 KR^{a,b}(X,Y) \stackrel{\cong}{\to} KR^{a+1,b+1}(X,Y),
\end{align}
and therefore one may define the positive K-groups as
\begin{align}
KR^{b}=KR^{0,b}.
\end{align}
Moreover we deduce natural isomorphisms
\begin{align}
KR^{a,b}\cong KR^{b-a} \cong \widetilde{KR}^0(S^{a,b}) .
\end{align}

Several important facts of the $KR$-theory need to be highlighted.
For spaces with trivial involution, the $KR$ groups simply become the real $KO$ groups. Hence for the fixed
point set of the involution $X^\tau$ we have: $KR^*(X^\tau)=KO^*(X^\tau)$. Moreover we have:
\begin{align} \label{isomorphism spheres}
\widetilde{KR}^{j}(S^{a,b}) \cong KO^{j-a+b}.
\end{align}

The $KR$-theory has period 8 by Bott periodicity of the orthonormal group $O$:
\begin{align}
KR^*(X,Y) \stackrel{\cong}{\to} KR^{*-8}(X,Y).
\end{align}

Whenever the complex anti-linear involution $J$ that lifts $\tau$ squares to $-1$ ($J^2=-1$), we call the vector
bundle quaternionic \cite{Gomi-quaternionic}. The Grothendieck group of quaternionic vector bundles over the involution space
is denoted by $KQ(X)$ and we may define the $KQ$ groups following the procedure described above: $KQ^{b-a}(X)= KQ^{a,b}(X)$.

For spaces with trivial involution the $KQ$-groups become the symplectic $KSp$-groups \cite{Dupont}. Hence for the fixed point set
of the involution we have  
\begin{align}
KQ^*(X^\tau)\cong KSp^*(X^\tau).
\end{align}

More importantly, the Bott periodicity implies that there is a canonical isomorphism \cite[\S 3.6]{Mishchenko}
\begin{align}
KR^{*+4} \cong KQ^{*}.
\end{align}

In Table \ref{KO and KSp groups} we recall the $KO$ and $KSp$ groups of a point.
\begin{table}
\begin{tabular}{||c||c|c|c|c|c|c|c|c||} 
\hline
* & 0 & 1 & 2 & 3 & 4 & 5 & 6 & 7 \\ 
\hline
$KO^*$ & $\mathbb{Z}$ & 0 & 0 & 0 & $\mathbb{Z}$ & 0 & $\mathbb{Z}_2$ & $\mathbb{Z}_2$  \\
\hline
$KSp^*$ & $\mathbb{Z}$ & 0 & $\mathbb{Z}_2$ & $\mathbb{Z}_2$ & $\mathbb{Z}$   & 0 & 0 & 0 \\
\hline
\end{tabular} \label{homotopy groups of KO}
\caption{$KO$ and $KSp$ groups of a point. } 
\label{KO and KSp groups}
\end{table}

\subsection{2-fold rotation composed with time reversal}

As it was pointed before, the 2-fold symmetries $C_{2w}$ and $C_{2 {\bf d}}$ have different topological properties. Let us start
with the rotation around a coordinate axis and later we will present the rotation around a diagonal.

\begin{table*}
\begin{tabular}{||c|c|c||c|c|c|c|c|c|c|c||} 
\hline
\multicolumn{10}{||c||}{{K-theory invariants of the equivariant cells $(C,\partial C)$  in 3-dimensional torus}}\\
\hline
Symmetry & Square  & K-group &  $k_x$-axis & $k_y$-axis & $k_z$-axis & $k_x=0$ & $k_y=0$ & $k_z=0$ & Bulk \\
\hline
\multirow{2}{*}{$\mathbb{T}$} & \multirow{2}{*}{-1} &  
 \multirow{2}{*}{$\widetilde{KR}^4$}  &  $\widetilde{KR}^4(S^{0,1})$ & $\widetilde{KR}^4(S^{0,1})$  &   $\widetilde{KR}^4(S^{0,1})$  &$\widetilde{KR}^4(S^{0,2})$  &  $\widetilde{KR}^4(S^{0,2})$  &  $\widetilde{KR}^4(S^{0,2})$   &  $\widetilde{KR}^4(S^{0,3})$\\
 & &  & 0 & 0&0&$\mathbb{Z}_2$&$\mathbb{Z}_2$&$\mathbb{Z}_2$&$\mathbb{Z}_2$ \\
\hline
\multirow{2}{*}{$C_{2z} \mathbb{T}$} & \multirow{2}{*}{1} &  
\multirow{2}{*}{$\widetilde{KR}^0$}  & $\widetilde{KR}^0(S^{1,0})$ & $\widetilde{KR}^0(S^{1,0})$  &   $\widetilde{KR}^0(S^{0,1})$  &$\widetilde{KR}^0(S^{1,1})$  &  $\widetilde{KR}^0(S^{1,1})$  &  $\widetilde{KR}^0(S^{2,0})$   &  $\widetilde{KR}^0(S^{2,1})$\\
 & &  &      $\mathbb{Z}_2$ & $\mathbb{Z}_2$ & 0 & $\mathbb{Z}$ & $\mathbb{Z}$ & $\mathbb{Z}_2$ & $\mathbb{Z}_2$ \\
 \hline
 \multirow{2}{*}{$C_{2( 1\bar{1}0 )} \mathbb{T}$} & \multirow{2}{*}{1} &  
\multirow{2}{*}{$\widetilde{KR}^0$}  & \multicolumn{2}{|c|}{$\widetilde{K}^0(S^{1})$} & $\widetilde{KR}^0(S^{1,0})$   &    \multicolumn{2}{|c|}{$\widetilde{K}^{0}(S^{2})$}  &  $\widetilde{KR}^{0}(S^{1,1})$   &  $\widetilde{KR}^{0}(S^{2,1})$\\
  & & &      \multicolumn{2}{|c|}{0} & $\mathbb{Z}_2$ & \multicolumn{2}{|c|}{$\mathbb{Z}$} & $\mathbb{Z}$ & $\mathbb{Z}_2$ \\
 \hline
\end{tabular}

\caption{Topological invariants associated to the symmetries $\mathbb{T}$,  $C_{2z} \mathbb{T}$ and $C_{2 {\bf d}}\mathbb{T}$ for ${\bf d}=(1\bar{1}0)$. Here we have calculated
the relative $K$-groups for each of the involutions in momentum space of the symmetries of first column. The second column describes
whether the operator can be thought of as real (square $1$) or quaternionic (square $-1$), and the third column
describes the appropriate K-theory groups that apply to the symmetry: $KR^0$ in the real case and $KR^4$ in the quaternionic case.
 The rest of the columns
are the topological invariants ($K$-groups) associated to each cell on the 3D torus (the 1D cells are the coordinate axis, the 2D cells are the 
coordinate planes and the 3D cell is the bulk). The K-theory invariants are the ${KR}^0$ groups for the relative pairs $(C, \partial C)$
whenever $C$ is an equivariant cell. Note that the $C_{2( 1\bar{1}0)}$ symmetry swaps the $k_x$ and $k_y$ axis and the 
$k_x=0$ and $k_y=0$ planes; hence the relative K-groups on those subspaces incorporate both lines and both planes respectively.
The first two rows depend on the stable splittings of Eqns. \eqref{stable splitting T} and \eqref{stable splitting C2zT}
respectively, and the third row 
depends on the decomposition in equivariant cells presented in \eqref{decomposition C2dT}.} 
\label{Table KR-theory invariants}
\end{table*}

\subsubsection{Symmetries $\mathbb{T}$ and $C_{2w}\mathbb{T}$}

Assume that the rotation is around the $z$-axis $C_{2z}$ and take the induced action in momentum space
\begin{align}
\mathbb{T}(k_x,k_y,k_z) &= (-k_x,-k_y,-k_z),\\
 C_{2z}  \mathbb{T} (k_x,k_y,k_z) &= (k_x,k_y,-k_z).
\end{align}

Denote by $S^1:=S^{1,0}$ the circle with trivial involution and $\mathbb{S}^1:=S^{0,1}$ the circle with
the sign involution. Note that the induced action on the Brillouin zone become
\begin{align}
\mathbb{T} \  : & \  \mathbb{S}^1 \times \mathbb{S}^1 \times \mathbb{S}^1\\
C_{2z} \mathbb{T} : & \  S^1 \times S^1 \times \mathbb{S}^1.
\end{align}

In order to calculate the K-theory groups we need to use the following stable splitting \cite[Thm. 11.8]{Freed-Moore}.
Consider the action of the group $(\mathbb{Z}_2)^3$ on $(S^1)^3$ given by multiplication of $-1$
on each coordinate. Then $(S^1)^3$ is $(\mathbb{Z}_2)^3$-equivariantly stably homotopy equivalent to a wedge of spheres:
\begin{align} \label{stable splitting 3d torus}
(S^1)^3 \sim S^3 \vee S^2 \vee S^2 \vee S^2 \vee S^1 \vee S^1 \vee S^1.
\end{align}

For the $\mathbb{Z}_2$-action given by both $\mathbb{T}$ and $I$ we have the equivariant stable equivalence:
 \begin{align} \label{stable splitting T} 
(\mathbb{S}^1)^3 \sim S^{0,3} \vee (S^{0,2})^{\vee 3} \vee (S^{0,1})^{\vee 3},
\end{align}
and for the $\mathbb{Z}_2$-action given by $C_{2z} \mathbb{T}$ we have the equivariant stable splitting:
 \begin{align}  \label{stable splitting C2zT} 
S^1 \times S^1 \times \mathbb{S}^1 \sim S^{2,1} \vee S^{2,0} \vee (S^{1,1})^{\vee 2} \vee S^{0,1} \vee (S^{1,0})^{\vee 2}.
\end{align}

For the antiunitary symmetries $\mathbb{T}$ and $C_{2z} \mathbb{T}$ the relevant K-theory groups are $KR^4$ and
$KR^0$ respectively since $\mathbb{T}^2=-1$ while $(C_{2z} \mathbb{T})^2=1$. Applying the isomorphism of Eqn.
\eqref{isomorphism spheres} together with the stable splitting described above, 
we obtain the following decomposition for the K-theory groups associated to the symmetry $C_{2z} \mathbb{T}$:
\begin{align}
KR^0&(S^1 \times S^1 \times \mathbb{S}^1) \\
  \cong & \widetilde{KR}^0(S^{2,1}) \oplus \widetilde{KR}^0(S^{1,1})^{\oplus 2} \oplus \widetilde{KR}^0(S^{2,0}) \\
 & \oplus \widetilde{KR}^0(S^{1,0})^{\oplus 2} \oplus \widetilde{KR}^0(S^{0,1}) \oplus {KR}^0(*)\\
 \cong & \mathbb{Z}_2 \oplus \mathbb{Z}^{\oplus 2} \oplus \mathbb{Z}_2 \oplus  (\mathbb{Z}_2)^{\oplus 2}  \oplus  0  \oplus \mathbb{Z}.
\end{align}

Incorporating the homotopy groups of $KO$ given in Table \ref{KO and KSp groups},
we obtain the K-theory groups summarized in Table \ref{Table KR-theory invariants}.

\subsubsection{Symmetry $C_{2{\bf d}}\mathbb{T}$}

Assume that the rotation $C_{2{\bf d}}$ is around ${\bf d} = ( 1\bar{1}0)$ and take the induced action in momentum space
\begin{align}
 C_{2{\bf d}}  \mathbb{T} (k_x,k_y,k_z) &= (-k_y,-k_x,k_z).
\end{align}

The appropriate K-theory is the real K-theory $KR^0$ since $ (C_{2{\bf d}}  \mathbb{T})^2=1$.
Decomposing the torus $(S^1)^3 \cong [0,2\pi)^3$ in equivariant cells and calculating the relative ${KR}^0$ groups
 we obtain the following information:
\begin{align}   \label{decomposition C2dT}
\begin{tabular}{|c|c|c|}
\hline
Cell=$C$  & ${KR}^0(C, \partial C)$ & Invariant\\
\hline \hline
$(0,2\pi)^3$ & $\widetilde{KR}^0(S^{2,1})$ & $\mathbb{Z}_2$ \\
\hline
$(0,2\pi)^2 \times \{0\}$ & $\widetilde{KR}^0(S^{1,1})$ & $\mathbb{Z}$  \\
\hline
$\{0\} \times (0,2\pi)^2 \cup $ & \multirow{2}{*}{$\widetilde{K}^0(S^{2})$} & \multirow{2}{*}{$\mathbb{Z}$}  \\
$(0,2\pi) \times \{0\} \times (0,2\pi)$ & & \\
\hline
$\{0\} \times \{0\} \times (0,2\pi)$ & $\widetilde{KR}^0(S^{1,0})$ & $\mathbb{Z}_2$  \\
\hline
$ (0,2\pi) \times \{0\} \times \{0\} \cup $ & \multirow{2}{*}{$\widetilde{K}^0(S^{1})$} & \multirow{2}{*}{0}  \\
$\{0\} \times (0,2\pi) \times \{0\} $ & & \\
\hline
\end{tabular} 
\end{align}

Note that the projection into the $k_z$ coordinate has a section, namely the composition of the following maps
is the identity:
\begin{align}
S^1 \to & (S^1)^3 \stackrel{\pi_z}{\longrightarrow} S^1\\
k_z \mapsto &(0,0,k_z) \mapsto k_z,
\end{align}
and moreover the maps are $C_{2{\bf d}}\mathbb{T}$-equivariant. This implies that the pullback map $\pi_z^*$ 
in K-theory is injective, and therefore the K-theory of $(S^1)^3$ subject to the $C_{2{\bf d}}\mathbb{T}$
symmetry is precisely isomorphic to the groups presented above:
\begin{align} \label{K-theory C2dT}
\widetilde{KR}^0_{C_{2{\bf d}}\mathbb{T}}((S^1)^3 ) \cong \mathbb{Z}_2 \oplus \mathbb{Z} \oplus \mathbb{Z} \oplus \mathbb{Z}_2.
\end{align}

The first one is a $\mathbb{Z}_2$  bulk invariant that will be studied in the next section,
the second and third $\mathbb{Z}$ invariants come from Chern classes, and the last $\mathbb{Z}_2$
invariant comes from the real K-theory of the $k_z$-axis.

Now let us see what happens with the restriction map to the fixed points of the symmetry
 $C_{2{\bf d}}\mathbb{T}$. On fixed points the operator $C_{2{\bf d}}\mathbb{T}$
 is trivial, and on the states it acts as complex conjugation since it squares to 1. The appropriate
 K-theory group on the fixed point set is then $KO^0$.
The $ C_{2{\bf d}}  \mathbb{T}$-fixed points define the 2D torus
\begin{align}
\left((S^1)^3 \right)^{C_{2{\bf d}}  \mathbb{T}} = \{(k,-k,k_z) : k,k_z \in S^1 \}
\end{align}

and the restriction map to the fixed points induce the map
\begin{align}  \label{restriction of KR to KO}
\widetilde{KR}^0_{C_{2{\bf d}}\mathbb{T}}((S^1)^3 ) \to \widetilde{KO}^0\left(\left((S^1)^3 \right)^{C_{2{\bf d}}  \mathbb{T}}\right).
\end{align}
By the stable splitting mentioned above, the right hand side of \eqref{restriction of KR to KO}
is isomorphic to 
\begin{align}
\widetilde{KO}^0(S^2) \oplus \widetilde{KO}^0(S^1)^{\oplus 2} \cong \mathbb{Z}_2^{\oplus 3},
\end{align} and the different terms of \eqref{decomposition C2dT}
restrict as follows: 
\begin{align}
\widetilde{KR}^0(S^{2,1}) \stackrel{\cong}{\to}&  \widetilde{KO}^0(S^2) & \mathbb{Z}_2 & \stackrel{\cong}{\to}\mathbb{Z}_2, \label{iso KR=(S2,1) to KO0(S2)} \\
\widetilde{KR}^0(S^{1,1}) \twoheadrightarrow&  \widetilde{KO}^0(S^1) & \mathbb{Z} & \twoheadrightarrow \mathbb{Z}_2, \label{map KR=(S,1) to KO0(S1)} \\
\widetilde{KR}^0(S^{1,0}) \stackrel{\cong}{\to}&  \widetilde{KO}^0(S^1) & \mathbb{Z}_2 & \stackrel{\cong}{\to}\mathbb{Z}_2. \label{iso KR=(S1,0) to KO0(S1)}
\end{align}
The isomorphism of \eqref{iso KR=(S2,1) to KO0(S2)} it is shown in Eqn. \eqref{K-theory invariant S2,1} in the next section
and restricts the Kane-Mele bulk invariant in $S^{2,1}$ to the second Stiefel-Whitney class of the sphere $S^2$, the homomorphism of \eqref{map KR=(S,1) to KO0(S1)} restricts the Chern class of $S^{1,1}$ to the first Stiefel-Whitney class of the circle $S^{1,0}$, and 
the isomorphism \eqref{iso KR=(S1,0) to KO0(S1)} is simply given by the first Stiefel-Whitney class of the bundle on the $k_z$-axis.

The restriction homomorphism is then:
\begin{align}  \label{restriction of KR to KO complete}
\widetilde{KR}^0_{C_{2{\bf d}}\mathbb{T}}((S^1)^3 ) \to & \widetilde{KO}^0\left(\left((S^1)^3 \right)^{C_{2{\bf d}}  \mathbb{T}}\right)\\
\mathbb{Z}_2 \oplus \mathbb{Z} \oplus \mathbb{Z} \oplus \mathbb{Z}_2 \to & \mathbb{Z}_2 \oplus \mathbb{Z}_2 \oplus \mathbb{Z}_2 \\
(a,b,c,d) \mapsto & (a,b,d).
\end{align}

Notice that the restriction homomorphism is surjective.

\subsection{Bulk invariant} \label{section bulk invariant}

From Table \ref{Table KR-theory invariants} we see that the bulk invariant is $\mathbb{Z}_2$ for systems with either $\mathbb{T}$, $C_{2w} \mathbb{T}$ or
 $C_{2{\bf d}}  \mathbb{T}$  symmetry.
In the case of time reversal symmetry this invariant is known as the Kane-Mele invariant and comes from the K-theory group
\begin{align} \label{K-theory T}
\widetilde{KR}^4(S^{0,3}) \cong \mathbb{Z}_2.
\end{align}
In the case of the composition of a two-fold rotation with time reversal ($C_{2w} \mathbb{T}$ or
 $C_{2{\bf d}}  \mathbb{T}$ ) the invariant comes from the K-theory group
\begin{align} \label{K-theory C2T}
\widetilde{KR}^0(S^{2,1}) \cong \mathbb{Z}_2
\end{align}
 but it has no proper name. Since
the K-theory construction is similar in all cases, we will denote both invariants as the {\it K-theory Kane-Mele invariants}. 
Here we note that the relative terms $(C,\partial C)$ are equivalent for both systems $C_{2w} \mathbb{T}$ and
 $C_{2{\bf d}}  \mathbb{T}$ whenever $C$ is the bulk. In both cases the relative term is equivalent to 
 the relative term $(B^{2,1}, \partial B^{2,1}) \simeq S^{2,1}$. Hence
 in this section we will focus on the symmetry $C_{2z} \mathbb{T}$; the results apply to both symmetries.

Explicit generators for both groups of Eqns.  \eqref{K-theory T} and \eqref{K-theory C2T} can be obtained
from a general construction incorporating the symmetries $\mathbb{T}$, $C_{2z}$ and $I$.
Consider the 3-dimensional sphere 
\begin{align}
S^3 = \{ (t, k_x,k_y,k_z) | t^2 + k_x^2+k_y^2+k_z^2 = 1\}
\end{align}
and take the actions
\begin{align}
\mathbb{T}  (t, k_x,k_y,k_z) &=  (t,- k_x,-k_y,-k_z)\\
C_{2z} \mathbb{T} (t, k_x,k_y,k_z) &=  (t, k_x,k_y,-k_z).
\end{align}
The sphere should be understood as the one-point compactification of $B^{0,3}$ for the time reversal symmetry and
of $B^{2,1}$ for $C_{2z} \mathbb{T}$. Since any complex vector bundle over $S^3$ trivializes, we may consider
the trivial rank $2$ complex vector bundle
\begin{align} \label{bundle E}
E : = S^3 \times \mathbb{C}^2.
\end{align}
The action on the fibers will be defined with the help of the parametrized matrices
\begin{align}
F : S^3 \to & SU(2)\\
 (t, {\bf k}) \mapsto & \left(
 \begin{matrix}
 t+ik_x & k_z+ik_y \\
 -k_z + ik_y & t - ik_x 
 \end{matrix}
 \right)
\end{align}
by the following equations:
\begin{align}
\mathbb{T} ((t, {\bf k}),{\Phi}) & = ((t, -{\bf k}), F(t,{\bf k}) \mathbb{J} {\Phi}), \\
I ((t, {\bf k}),{\Phi}) & = ((t, -{\bf k}), F(t,{\bf k}) {\Phi}), \label{action I}\\ 
\mathbb{T} I ((t, {\bf k}),{\Phi}) & = ((t, {\bf k}), \mathbb{J}  \Phi), \\
C_{2z} ((t, {\bf k}),{\Phi}) & = ((t, -k_x,-k_y,k_z),i \sigma_2{\Phi}), \\
C_{2z} \mathbb{T}  ((t, {\bf k}),{\Phi}) & = ((t, k_x,k_y,-k_z),i \sigma_2F(t,{\bf k}) \mathbb{J}{\Phi}), \label{action C2zT}
\end{align}
where $\mathbb{J}= i \sigma_2  \mathbb{K}$, $\sigma_j$ are the Pauli matrices and $\mathbb{K}$ denotes complex conjugation.
The action is well defined because of the following equalities:
\begin{align}
\mathbb{K} F(t,{\bf k})=& F(t,C_{2z} {\bf k}) \mathbb{K}\\
i \sigma_2 F(t,{\bf k})=& F(t,C_{2z} {\bf k}) i \sigma_2\\
 F(t,{\bf k})^{-1}=& F(t,-{\bf k}).
\end{align}

It is well known that the isomorphism class of $E$ together with the $\mathbb{T}$ symmetry generates the group $KR^4(S^{0,3})$:
\begin{align}
\langle  [E] \rangle \cong KR^4(S^{0,3}) \cong \mathbb{Z}_2.
\end{align}
 This can be seen
from the eigenvalues of $I$ on its fixed points $(\pm 1, {\bf 0})$:  the two eigenvalues are $+1$ over $( 1, {\bf 0})$ and both are $-1$ on 
$( -1, {\bf 0})$. Then the Fu-Kane-Mele formula \cite{Fu-Kane-Mele} for the fixed points of the inversion action implies the claim.

To show that the isomorphism class of $E$ together with the $\mathbb{T} C_{2z}$ symmetry generates the group 
$KR^0(S^{2,1})$ we need to further elaborate. Note that restricting to the fixed points of the operator $\mathbb{T} C_{2z}$
we obtain a canonical isomorphism:
\begin{align} \label{isomorphims KR(S21) to KO(S2)}
\widetilde{KR}^0(S^{2,1}) \stackrel{\cong}{\to} \widetilde{KO}^0(S^{2}).
\end{align}
On the one side, the $\mathbb{K}$-equivariant maps to $BU$ provide the $KR$ groups:
\begin{align}
\widetilde{KR}^0(S^{2,1})  \cong \pi_0 \left( \mathrm{Map}(S^{2,1},BU)^{\mathbb{K}} \right), 
\end{align}
while on the other, its fixed point maps provide the $KO$ groups:
\begin{align}
\widetilde{KO}^0(S^{2})  \cong \pi_0 \left( \mathrm{Map}(S^{2},BO) \right). 
\end{align}
Here we are using the fact that the $\mathbb{K}$-fixed points of the complex Grassmannians are the real Grassmannians:
\begin{align}
BU^{\mathbb{K}} = BO.
\end{align}
Since $\pi_3(BU)=0$, any map from $S^2$ to $BO$ lifts to a map from $S^{2,1}$ to $BU$, and therefore
the isomorphism of Eqn. \eqref{isomorphims KR(S21) to KO(S2)} follows.

The $\mathbb{T}C_{2z}$-fixed points of $E$ consists of the real bundle over the sphere 
\begin{align}
S^2 = \{ (t,k_x,k_y,0) \in S^{2,1} \}
\end{align}
whose vectors satisfy the equation
\begin{align}
\left(
\begin{matrix}
-k_y & -k_x+it \\
k_x+it & -k_y
\end{matrix} \right) \mathbb{K} \Phi = \Phi.
\end{align}
On the equator, that is whenever $t=0$, we observe that the real matrix provides the generator of $\pi_1(O(2)) \cong \mathbb{Z}$
which induces the generator of $\pi_1(O) \cong  \mathbb{Z}_2$. This implies that the isomorphism class of the fixed point real bundle $E^{\mathbb{T}C_{2z}}$
is the generator of $\widetilde{KO}^0(S^2)$ and therefore the isomorphism class of $E$ generates  $\widetilde{KR}^0(S^{2,1})$:
\begin{align}  \label{K-theory invariant S2,1}
\widetilde{KR}^0(S^{2,1}) \cong \langle [E] \rangle \stackrel{\cong}{\to} \langle [E^{\mathbb{T}C_{2z}}] \rangle \cong \widetilde{KO}^0(S^2).
\end{align}

Since the symmetries $C_{2z} \mathbb{T}$ and $C_{2{\bf d}}  \mathbb{T}$ are equivalent on the relative term $(C, \partial C)$
whenever $C$ is the bulk, we see that the isomorphism class of $E$ generate the $\mathbb{Z}_2$ invariant of the bulk in both cases.
It should be also noted that the K-theory Kane-Mele invariant restricts to the second Stiefel-Whitney class
of the sphere $S^2$.

\subsection{Band inversion} \label{section band inversion}

The explicit bundle $E$ defined in Eqn. \eqref{bundle E}, which generates the K-theory groups
$\widetilde{KR}^0(S^{0,3})$  and $\widetilde{KR}^0(S^{2,1})$ associated to the bulk invariant
of the symmetries $\mathbb{T}$  and $C_{2z} \mathbb{T}$ respectively, incorporates the information
required to describe the behavior of energy bands that undergo te change of orbital type known as  {\it band inversion}.
This happens whenever two energy bands change of orbital type around a high symmetry point. 
This relation was noticed by the authors in \cite{Gonzalez-Pinilla-Uribe} for the symmetry $\mathbb{T}$; here we show that
the same relation also holds for the symmetry $C_{2z}\mathbb{T}$. 

Define the rank 2 complex vector bundle $L \subset S^3 \times (\mathbb{C}^2)^2$ isomorphic to $E$ by the following equation:
\begin{align}
L= \left\{ \left[ (t, {\bf k}),   \left( \left(\tfrac{1+t}{2}\right)^{\frac{1}{2}} u,  \left(\tfrac{1-t}{2}\right)^{\frac{1}{2}} \tfrac{1}{|{\bf k|}}F(0,{\bf k}) u \right) \right] : u \in \mathbb{C}^2 \right\},
\end{align}
and define the actions of $I$ and $C_{2z}\mathbb{T}$ by the following formulas:
\begin{align}
I  \cdot \left( (t , {\bf k}), (u,v) \right) = & \left( (t , -{\bf k}), (u,-v) \right)\\
C_{2z}\mathbb{T}  \cdot \left( (t , {\bf k}), (u,v) \right) =  &\left( (t , k_x,k_y,-k_z, (\mathbb{K}u,-\mathbb{K}v) \right)
\end{align}
where $\mathbb{K}$ denotes complex conjugation.

The bundle $L$ models the change of orbital type around a high symmetry point. If we take the first coordinate $u$ 
as the projection into the first orbital, and the second coordinate $v$ as the projection into the second orbital,
we see that on the point $(1, {\bf 0})$ the vectors are completely on the first orbital while on $(-1, {\bf 0})$
the vectors are completely on the second. The states on the vector bundle $L$ change of orbital type from
$(1, {\bf 0})$ to $(-1, {\bf 0})$. The eigenvalues of the inversion operator change accordingly. They are both $+1$ on
the first point and both $-1$ on the second point.

To witness the band inversion on a system preserving both inversion and the $C_{2z} \mathbb{T}$ symmetry, we have constructed
a $4 \times 4$ Hamiltonian with the following form:
\begin{align} \label{local Hamiltonian}
H({\bf k}) = &M({\bf k}) \tau_3 \sigma_0 + A({\bf k}) \tau_1 \sigma_3 + B({\bf k}) \tau_2 \sigma_0 \\
&+ C({\bf k}) \tau_1 \sigma_1 + D({\bf k}) \tau_1 \sigma_2 +W({\bf k}) \tau_0 \sigma_1.   
\end{align}
Here $\tau_i$ and $\sigma_j$ are Pauli matrices in orbital and spin coordinates respectively, and the functions are:
\begin{align}
M({\bf k})= & D_1 -m_1k_z^2-n_1(k_x^2+k_y^2)\\
A({\bf k}) = &D_2k_z +E_2k_z^3 + F_2k_z(k_x^2+k_y^2)\\
B({\bf k}) = &D_3k_xk_yk_z\\
C({\bf k}) = &(D_5k_z^2+D_6)k_x\\
D({\bf k}) = &(D_7k_z^2+D_8)k_y\\
W({\bf k}) = &D_4((k_x+ik_y)^2+k_z^2).  
\end{align}
The Hamiltonian preserves both inversion and $C_{2z} \mathbb{T}$ symmetries with matrices:
\begin{align}
I = \tau_3\sigma_0 \ \ \mathrm{and} \ \ C_{2z}\mathbb{T}=i\tau_0\sigma_1 \mathbb{K},
\end{align}
while it breaks time reversal symmetry $ \mathbb{T}=i \tau_0\sigma_2 \mathbb{K}$ because
$W({\bf k})=W(-{\bf k})$ and the operators $i \tau_0\sigma_2 \mathbb{K}$ and $ \tau_0 \sigma_1$ anticommute.

\begin{figure}
\includegraphics[width=8.8cm]{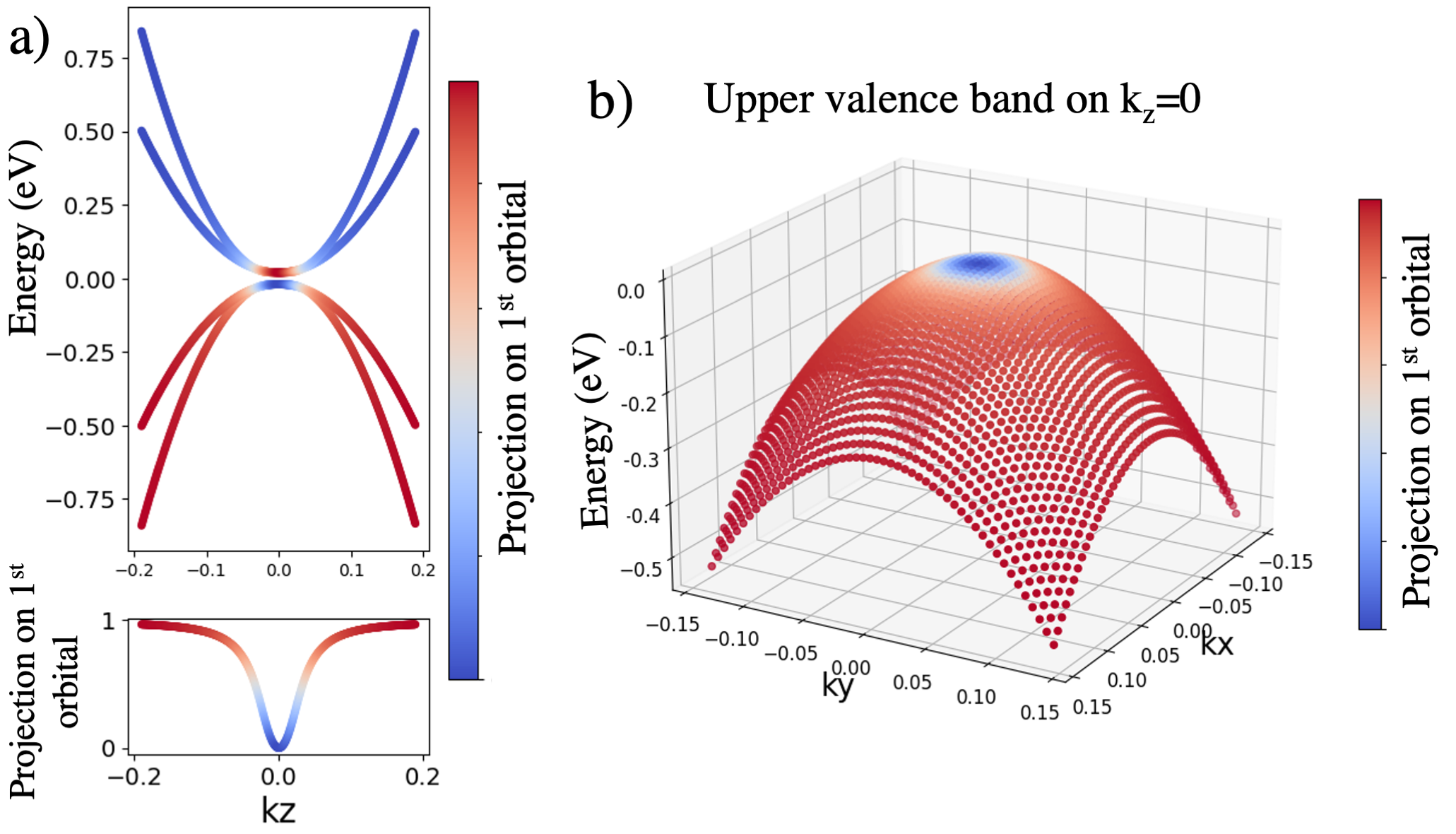}
\caption{Energy bands of the local Hamiltonian of Eqn. \eqref{local Hamiltonian} preserving both
inversion and $C_{2z}\mathbb{T}$ but breaking time reversal. a) Energy bands along the $k_z$-axis with the projection on the
first orbital of the upper valence band. b) Upper valence band on the $k_z=0$ with its projection on the first
orbital. Note the change of orbital type along the energy bands.}
\label{Figure local Hamiltonian}
\end{figure}

Setting up the constants to the following values:
\begin{align}
D_1=&0.02,   
m_1=n_1=18, 
D_2=D_6=D_8=1\\
&E_2=F_2=D_3=D_4=D_5=D_7=5
\end{align}
we obtain the energy bands that can be observed in Fig. \ref{Figure local Hamiltonian}. 
If the Fermi level is set to be zero we observe a small energy gap. The two valence bands
meet at the origin and they change of orbital type, being completely on the first orbital in the origin, 
and almost completely on the second whenever $|{\bf k}| > 0.2$. This change of orbital type 
of two valence bands around a high symmetry point encodes the K-theory Kane-Mele bulk invariant introduced in
Eqn. \eqref{K-theory invariant S2,1} 
that the symmetry  $C_{2z} \mathbb{T}$ protects.

{\subsection{Chern-Simons axion coupling}

In several works \cite{Equivalent_topological_invariants_of_topological_insulators, Inversion_symmetric_topological_insulators, Symmetry_representation_approach_to_topological_invariants}
 it has been shown that, under certain hypothesis, the parity of the degree of the sewing
matrices of a symmetry of the system recovers the Chern-Simons axion coupling \eqref{Chern-Simons axion coupling}
for axion insulators.
To be more precise, take $U$ a symmetry of the system acting on the crystal as $({\bf r},t) \mapsto (O{\bf r} + {\bf t}, s_Ut)$
where $O$ is the point group symmetry, ${\bf t}$ is a translation and $s_U= \pm 1$. 
Let $|\psi_{n,{\bf k}} \rangle$ be the eigenvectors of the Hamiltonian and assume that they can be defined
continuously on the whole torus $T^3$. Note that this can only be achieved whenever all the Chern numbers are zero along all planes
(the Chern classes are precisely the obstruction for the existence of global sections on rank $1$ complex bundles).
Since the existence of Chern insulators have been elusive, we may assume for the time being that this is the case for all crystals.

Define the sewing matrices $G(U)$ of the operator $U$ by the following equation:
\begin{align}
G(U)_{mn}({\bf k})= \langle \psi_{m, s_gO{\bf k}} | U | \psi_{n,{\bf k}} \rangle.
\end{align}
Whenever $s_U \mathrm{det}(O)=-1$, or in other words, whenever $O$ reverses orientation and time is  fixed or $O$ preserves orientation and time is reversed,
then
\begin{align}
2\frac{\theta}{2\pi}=\mathrm{deg}(G(U)) \  \mathrm{mod} \ 2 
\end{align}
where $\theta$ is the Chern-Simons axion coupling and $\mathrm{deg}(G(U))$ is the degree if the sewing matrices $G(U)$ which is defined
by the equation:
\begin{align}
\mathrm{deg}(G(U)) = \frac{1}{24\pi^2} \int_{\mathrm{BZ}} d^3k
\mathrm{Tr}(G(U)^{-1}dG(U))^3
\end{align}

This equation has been shown in \cite[Appendix D]{Symmetry_representation_approach_to_topological_invariants}
and requires the triviality of all Chern numbers over all surfaces as it was imposed above.

Take the bundle $E$ with the symmetry $C_{2z} \mathbb{T}$ defined in Eqn. \eqref{action C2zT}. Here
the bundle is defined over the 3D sphere and therefore all Chern numbers vanish. The sewing matrices
for the operator $C_{2z} \mathbb{T}$ become:
\begin{align}
G(C_{2z} \mathbb{T}) : S^3& \to SU(2) \\
(t, {\bf k}) & \mapsto i \sigma_2 F(t,{\bf k})
\end{align}
and its degree is $\pm 1$ (depending on the orientation of $SU(2)$). The parity of the degree
of the sewing matrices is nothing else but the K-theory Kane-Mele invariant of the bundle $E$ that we defined
in \eqref{K-theory invariant S2,1}. Hence the non-triviality of the Chern-Simons axion coupling on $E$ is equivalent to the non-triviality of the 
isomorphism class of $E$ in the K-theory group $\widetilde{KR}^0(S^{2,1}) \cong \mathbb{Z}_2$. 

We conclude that for the bundle $E$ subject to $C_{2z}\mathbb{T}$ defined in Eqn. \eqref{action C2zT}, the following
invariants are equivalent: the K-theory Kane-Mele invariant, the parity of the degree
of the sewing matrices $G(C_{2z}\mathbb{T})$ and the Chern-Simons axion coupling $\theta(E)$ of the system.

\subsection{Inversion} \label{section inversion}

If the system possessing the symmetry $C_{2}\mathbb{T}$ is moreover endowed with the inversion symmetry, we may read
the K-theory Kane-Mele invariant, or equivalently the Chern-Simons axion coupling, through the eigenvalues of the inversion
operator on the 8 time reversal invariant momenta (TRIMs). This is how it works.

Let us assume first that the only symmetry present is inversion.  We are interesting in determining the bulk
invariant in the $\mathbb{Z}_2$-equivariant complex K-theory group $\widetilde{K}_I^0((S^1)^3)$ where the inversion
operator $I$ generates de $\mathbb{Z}_2$-action. Later we will relate this bulk invariant
with the K-theory Kane-Mele invarint associated to the symmetries $\mathbb{T}$ and $C_2\mathbb{T}$.
 By the stable splitting of Eqn. \eqref{stable splitting 3d torus} we know that
 \begin{align}
 \widetilde{K}_I^0((S^1)^3) \cong \widetilde{K}_I^0(S^3) \oplus \widetilde{K}_I^0(S^2)^{\oplus 3} \oplus \widetilde{K}_I^0(S^1)^{\oplus 3}
 \end{align}
 where in all three spheres $S^k$ the inversion operator is acting by multiplication by $-1$ on the unit ball $B^k$ and
 we are taking the induced action on the spheres given by the homeomorphism $S^k \cong B^k / \partial B^k$.  The spheres
 described above can be seen as the following quotient spaces 
 \begin{align}
 S^3 \cong & (S^1)^3 / (k_x=0 \cup k_y=0 \cup k_z=0) \\
 S_i^2 \cong & (k_i=0) / (k_j-\mathrm{axis} \cup k_l-\mathrm{axis}), \ \ j \neq i \neq l \\
  S_i^1 \cong & k_i-\mathrm{axis}.
 \end{align}
In all seven spheres the point $(0,0,0)$ can be taken as the base point, and the other 7 TRIMs
belong each only to one sphere: $(\text{\textonehalf},\text{\textonehalf},\text{\textonehalf})$ in $S^3$, $(\text{\textonehalf},\text{\textonehalf},0)$ in $S^2_3$,
 $(\text{\textonehalf},0, \text{\textonehalf})$ in $S^2_2$, and so forth.
 
 Here we may assume
 that any complex vector bundle defined on $(S^1)^3$ has all $+1$ eigenvalues of $I$ on the point $(0,0,0)$,
 hence the topological information of the inversion operator will be localized on the other 7 TRIMs.
 
From the excision exact sequence of the pair $(B^3,S^2)$ with respect to the inversion action on $B^3$
(the center of $B^3$ is the point $(\text{\textonehalf},\text{\textonehalf},\text{\textonehalf})$), we
obtain the exact sequence:
\begin{align}
0 \to \widetilde{K}^0_I(S^3) \to& K^0_I(B^3) \to K^0(\mathbb{R}P^2) \\
0 \to \widetilde{K}^0_I(S^3) \to& \mathbb{Z}_+ \oplus \mathbb{Z}_- \to \mathbb{Z} \oplus \mathbb{Z}_2 \\
& (a,b) \mapsto (a+b,b),
\end{align}
where $\mathbb{Z}_\pm$ keeps track of the number of bands whose
inversion eigenvalue is $\pm1$ on the point $(\text{\textonehalf},\text{\textonehalf},\text{\textonehalf})$.
Here we have used the fact that $\widetilde{K}^0(\mathbb{R}P^2) \cong \mathbb{Z}_2$ and that its generator
can be obtained by the $-1$ action on the  fibers of the trivial bundle $\mathbb{C} \times S^2$.
Hence 
\begin{align} \label{inversion invariant}
\widetilde{K}^0_I(S^3) \cong 2 \mathbb{Z}_-
\end{align}
and therefore the $-1$ eigenvalues on the point $(\text{\textonehalf},\text{\textonehalf},\text{\textonehalf})$ on the sphere $S^3$
come in pairs.

Now note that the bundle $E$ of \eqref{bundle E} with the $C_{2z} \mathbb{T}$ and inversion actions defined in \eqref{action C2zT} and \eqref{action I} respectively, generates the group $\widetilde{K}^0_I(S^3)$ because
of the inversion eigenvalues on $(1,{\bf 0})$ and $(-1,{\bf 0})$, and also generates
$\widetilde{KR}^0(S^{2,1})$ by \eqref{K-theory invariant S2,1}. Hence, in order to detect the K-theory Kane-Mele invariant on the sphere $S^3$,
 it is enough to know that the number of $-1$ eigenvalues is congruent with $2$ modulo $4$.

The case of the 2D spheres $S^2_i$ is the following. The inversion operator acts by rotation and the exact sequence of the pair $(B^2,S^1)$ becomes:
\begin{align}
K^{-1}(S^1/I) \to \widetilde{K}^0_I(S_i^2) \to& K^0_I(B^2) \to K^0(S^1/I) \\
\mathbb{Z} \to \widetilde{K}^0_I(S_i^2) \to& \mathbb{Z}_+ \oplus \mathbb{Z}_- \to \mathbb{Z}  \\
& (a,b) \mapsto (a+b).
\end{align}
Therefore 
\begin{align}
 \widetilde{K}^0_I(S_i^2) \cong \mathbb{Z}\langle L_i^{\otimes 2} \rangle \oplus \mathbb{Z}_-\langle L_i \rangle
\end{align}
where the first component is generated by the line bundle $L_i^{\otimes 2}$ whose Chern number is 2  (and
therefore the action of $I$ on the fixed points of the inversion is trivial), and the second component counts the number
of $-1$ eigenvalues on the fixed point. If $L_i$ is the line bundle with $c_1(L_i)=1$,
we see that it has only one $-1$ eigenvalue and that it generates the group $\mathbb{Z}_-\langle L_i \rangle$.

We may choose a lift $\widetilde{K}^0_I(S_i^2) \to \widetilde{K}^0_I((S^1)^3)$ given by extending 
the bundle $L_i$ on the $k_i$-axis. Therefore the $-1$ eigenvalues on TRIMs are repeated twice and therefore
the contribution of the generic bundle  $n_i L_i^{\otimes h_i} \in \widetilde{K}^0_I(S_i^2)$ on the total number of $-1$ eigenvalues on TRIMs is $2n_i$ whenever $h_i$ is odd and $0$ whenever $h_i$ is even.

The bundles over the circles $S^1_i$ may have any eigenvalue on the fixed points. But whenever
they are lifted to $\widetilde{K}^0_I((S^1)^3)$ the eigenvalues are copied four times. Hence the number
of $-1$ eigenvalues coming from the circles is multiple of four.
Labeling the TRIMS with the eigenvalues of the inversion operator, we list
in Fig. \ref{fig TRIMS} a) our choice generators of the K-theory group $ \widetilde{K}_I^0((S^1)^3) $ which can be distinguished by these eigenvalues. Clearly
the line bundles $L^{\otimes 2m}$ cannot be distinguished since all their $I$-eigenvalues are positive.

 \begin{figure}
 	\includegraphics[width=8.8cm]{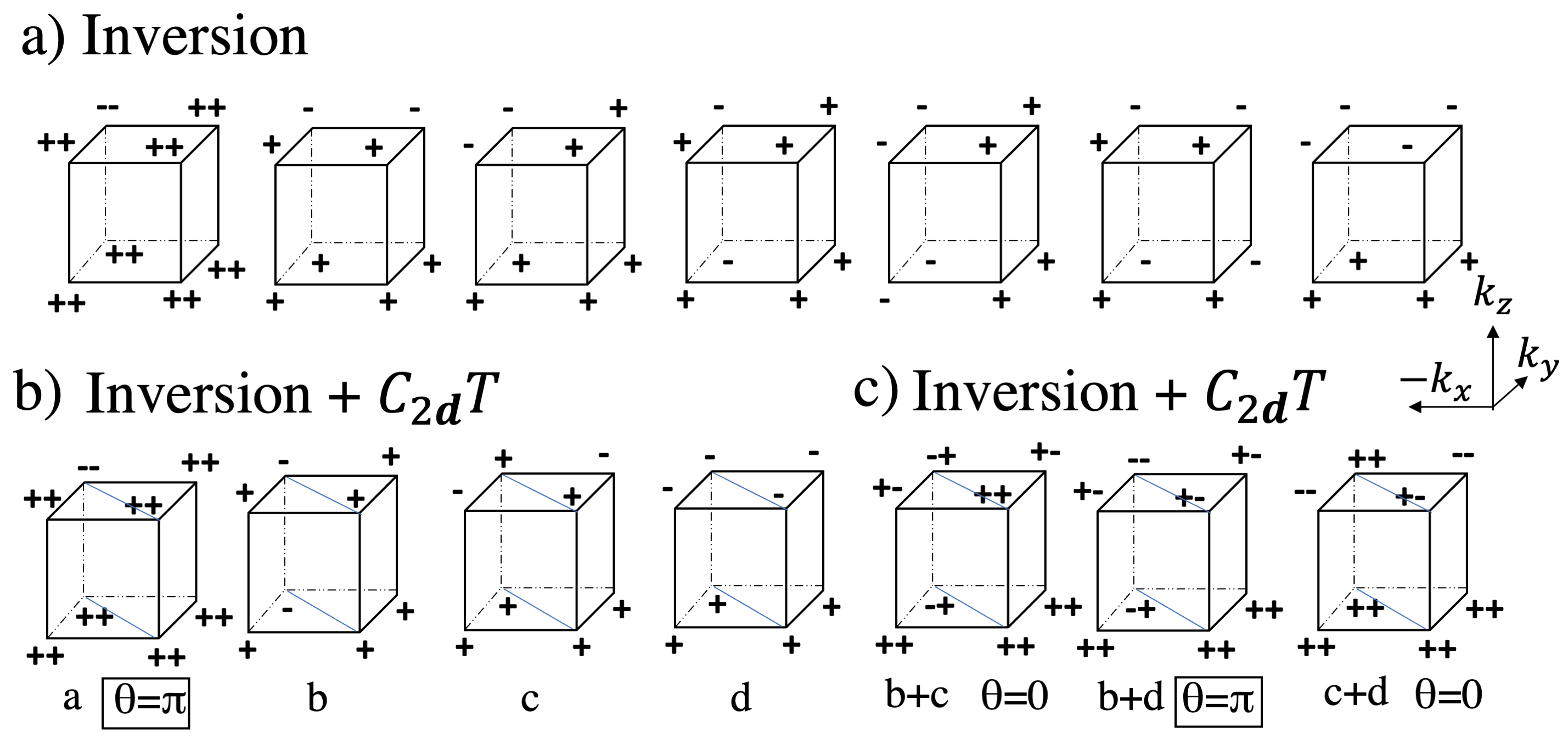}
 	\caption{Generators of the K-theory groups in the presence of inversion a) and $C_{2{\bf d}}\mathbb{T} $
 	and inversion b). The $\pm 1$ eigenvalues of the inversion operator are labeled on the eight TRIMs. In a)
 	the first one represents the bundle $E$ of Eqn. \eqref{bundle E}, the next three represent one band systems
 	with odd Chern numbers across the planes $k_x=0$, $k_y=0$ and $k_z=0$ respectively, and the last
 	three represent one band systems where the connection winds around the $k_x$, $k_y$ and $k_z$ axis respectively
 	with winding number 1. In b) the first one labeled with $a$ represents the bundle whose K-theory Kane-Mele invariant is non-trivial,
 	the second labeled with $b$ represents a one band system with odd Chern number across the plane $k_z=0$, the third one labeled with $c$ represents
 	a one band system with odd Chern numbers across $k_y=0$ and $k_x=0$ and the fourth labeled with $d$ represents a one band system
 	with winding number 1 along the $k_z$ axis. These generators are labeled according to Eqn. \eqref{CS formula general C2dT}. In c) all two different one band systems of b) are superposed. The
 	Chern-Simons axion coupling term is calculated as the parity of the number of pairs of negative $I$-eigenvalues on all TRIMs.}
 	\label{fig TRIMS}
 \end{figure}

If we denote $\xi_i$ the number of $-1$ eigenvalues on the $i$-th TRIMs, we may take
the congruence module four of these eight numbers. If we want to distinguish the Chern-Simons axion coupling term and  the K-theory Kane-Mele invariant we may take two approaches.

\subsubsection{Trivial Chern numbers} \label{section trivial Chern numbers}

If all the Chern numbers on all planes are trivial, then the K-theory Kane-Mele invariant 
{ is equivalent to the the Chern-Simons axion coupling term and}
is non-trivial whenever
\begin{align} \label{Z4 invariant}
\sum_{i \in \mathrm{TRIMs}} \xi_i \equiv \ 2 \ \mathrm{mod} \ 4. 
\end{align}
This follows from the fact that the bundle $E$, once pulled back to $(S^1)^3$ , 
has only two $-1$ eigenvalues on $(\text{\textonehalf},\text{\textonehalf},\text{\textonehalf})$ while having $+1$ eigenvalues on the
other seven TRIMs. This bundle is represented by $a$ in Fig. \ref{fig TRIMS} b).
This formula is pretty well known and has appeared in many works, among them \cite{Quantized_response_and_topology_of_magnetic_insulators, Inversion_symmetric_topological_insulators}.

\subsubsection{Non-trivial Chern numbers} \label{section Non-trivial Chern numbers}

Whenever the Chern numbers are not trivial, the K-theory Kane-Mele invariant and the Chern-Simons axion coupling term
may be different. The Chern-Simons axion coupling term, being a secondary characteristic class, incorporates
information of all bulk and surface invariants, while the K-theory Kane-Mele invariant is only a bulk invariant.

In this general setup, the Chern-Simons axion coupling term can be read
 from the parity of the total number of pairs of $-1$ eigenvalues that exist on each TRIMs by the following formula:
\begin{align} \label{formula with Chern classes}
\sum_{i \in \mathrm{TRIMs}} \left\lfloor \frac{\xi_i}{2} \right\rfloor \equiv 1 \ \mathrm{mod} \ 2.
\end{align} 
Here $ \lfloor x \rfloor$ denotes the greatest integer which is less or equal to $x$, and the formula
incorporates the information coming from the K-theory Kane-Mele invariant as well as information
coming from the coupling of two different Chern classes. 
Is it important to note that this formula has appeared previously in \cite{Band_Topology_and_Linking_Structure_of_Nodal_Line_Semimetals}
but for systems with underlying symmetry  $I \mathbb{T}$ and no SOC. In what follows we will show that it also works in systems
with inversion and $C_{2{\bf d}}\mathbb{T}$ symmetry. 

We have calculated the K-theory invariants for systems with $C_{2{\bf d}}\mathbb{T}$ in Eqn. \eqref{K-theory C2dT} and
in Fig. \ref{fig TRIMS} b) we have presented the choice of generators with the appropriate eigenvalues of the inversion operator on the TRIMs.

The Chern-Simons axion coupling term calculated on the generator of $\widetilde{KR}^0(S^{2,1})$ is non-trivial and it
 only incorporates information on the term $\mathcal{A}_\alpha \mathcal{A}_\beta
\mathcal{A}_\gamma$ because the curvatures $\mathcal{F}_{\alpha \gamma}$ are trivial, see \cite[Lem. 10.4]{Nakahara}

The rest of the generators of the K-theory group are composed of only one band, thus the Chern-Simons axion coupling term is
trivial on each one of them. But once these bands are taken in pairs of different bands they can define systems on which the 
Chern-Axion coupling term is non-trivial. 

Let us consider one band system which models the generator of the group $\widetilde{KR}^0(S^{1,1})$ which is defined
on the $k_z=0$ plane and copied along the $k_z$ direction. Let $\mathcal{A}_{x},\mathcal{A}_y$ and $\mathcal{F}_{xy}$ denote
its connection and curvature over the whole BZ and note that these are the only non-trivial components because all derivatives along
the $k_z$ axis are trivial. This band corresponds to system $b$ in Fig. \ref{fig TRIMS} b), its $I$-eigenvalues on $k_z=0$ and on $k_z=\pi$ are the same, and the
Chern number is $\frac{1}{2\pi}\int \mathcal{F}_{xy}dk_xdk_y=1$ whose parity can be read from the number of $-1$ eigenvalues on the $k_z$ plane.

Consider also the system of one band which models the generator of  $\widetilde{KR}^0(S^{0,1})$ along the $k_z$-axis. 
This is just a one band system on which the inversion operator acts multiplication by 0 on $k_z=0$ and by 1 on $k_z=\pi$
reproducing  the eigenvalues presented in system $d$ in Fig. \ref{fig TRIMS} b). Denote its connection $\overline{\mathcal{A}}_z$ and note that 
$\overline{\mathcal{A}}_x=0=\overline{\mathcal{A}}_y$ and moreover $\int \overline{\mathcal{A}}_z dk_z =2\pi$ 
(its winding number is 1).

The Chern-Simons axion coupling term of a system that incorporates the two previously defined bands (an abelian gauge) 
 is non-trivial and can be calculated as follows:
\begin{align}
\theta(\mathcal{A}\oplus\overline{\mathcal{A}}) = &  \frac{1}{4\pi} \int_{k_z}  \overline{\mathcal{A}}_z dk_z \int_{k_z=0} \mathcal{F}_{xy} dk_xdk_y\\ = &  \frac{1}{4\pi} 2\pi \cdot 2\pi = \pi
\end{align}
Note that the $I$-eigenvalues of the two band system previously defined are presented as $b+d$ in Fig. \ref{fig TRIMS} c) and
it has the important feature of having an odd number of TRIMs with both negative eigenvalues. 
This feature recognizes the non-triviality of the Chern-Simons axion coupling term and the formula presented in
Eqn. \eqref{formula with Chern classes} precisely distinguishes the parity of the number of TRIMs with both
negative eigenvalues.

The Chern-Simons axion coupling term of the other two groups of bands is trivial as can be distinguished in Fig. \ref{fig TRIMS} c) in systems labeled $b+c$ and $c+d$
since the number of TRIMs with  both negative $I$-eigenvalues is even.
We can bundle up the previous information in an explicit formula for the Chern-Simons axion coupling term in terms of generators in K-theory as follows:
\begin{align} \label{CS formula general C2dT}
\theta : \widetilde{KR}^0_{C_{2{\bf d}}\mathbb{T}}((S^1)^3) &\to \mathbb{Z}_2\\
\mathbb{Z}_2 \oplus \mathbb{Z} \oplus \mathbb{Z} \oplus \mathbb{Z}_2 & \to \mathbb{Z}_2 \\
(a,b,c,d)  & \mapsto   a + b\cdot d.  \label{CS formula general C2dT}
\end{align}
Hence we can conclude that the formula of Eqn. \eqref{formula with Chern classes} determines the Chern-Simons axion coupling term. 

Here we see the difference between the K-theory Kane-Mele invariant and the Chern-Simons axion coupling term:
the former is $a$ while the latter is $a + b \cdot d$. On systems without Chern numbers, both agree, but in general, they are different.
The difficulty arises when one wants to detect the Chern-Simons axion coupling term on a specific material whenever there is no inversion symmetry.
Without the eigenvalues of the inversion operator, it is difficult to determine the explicit values of $a, b, c$ and $d$. Therefore
the Chern-Simons axion coupling term is found by determining transport properties on surface or hinges as it is carried out on the
next chapter.

\subsection{Summary}

The symmetry $C_{2{\bf d}} \mathbb{T}$ with ${\bf d}=(1\bar10)$ endows a system with the K-theory invariants
$\mathbb{Z}_2 \oplus \mathbb{Z} \oplus \mathbb{Z} \oplus \mathbb{Z}_2$. The first $\mathbb{Z}_2$
comes from a 2-band system bulk invariant with trivial 2D curvatures and which is denoted the K-theory Kane-Mele invariant \S\ref{section bulk invariant};
systems with this invariant are expected to have band inversion \S \ref{section band inversion}. The first copy of the integers is generated by a
one band system whose Chern number across the $k_z=0$ plane is 1, and the second copy is generated by a one
band system whose Chern number across the $k_x=0$ plane is 1 and across the $k_y=0$ plane is -1. The last copy
of $\mathbb{Z}_2$ is generated by a one band system whose connection has winding number 1 along the $k_z$-axis 
\S \ref{section Non-trivial Chern numbers}.

The Chern-Simons axion coupling term is the parity of the sum of the K-theory Kane-Mele invariant with the multiplication
of the first integer with the second $\mathbb{Z}_2$ invariant as it is shown in  Eqn. \eqref{CS formula general C2dT}.

If inversion is moreover present, the Chern-Simons axion coupling term is equivalent to the parity of the number
\begin{align}
\sum_{i \in \mathrm{TRIMs}} \left\lfloor \frac{\xi_i}{2} \right\rfloor
\end{align}
where $\xi_i$ denotes the number of -1 eigenvalues of the inversion on the $i$-th TRIM, and $\lfloor x \rfloor$
is the largest integer less or equal to $x$.

Whenever all Chern numbers vanish, the K-theory Kane-Mele invariant is equivalent to the Chern-Simons axion coupling term \S \ref{section trivial Chern numbers}.
Therefore, if the Chern-Simons axion coupling term is non-trivial, band inversion should be expected on the valence bands.
If moreover, the system possesses inversion symmetry, the Chern-Simons axion coupling term can be read from the $\mathbb{Z}_4$-invariant
of Eqn. \eqref{Z4 invariant} which is given by the formula
\begin{align}
\sum_{i \in \mathrm{TRIMs}} \xi_i.
\end{align}
The system is trivial when the formula returns 0 modulo 4 and it is an axion insulator whenever the formula returns 2 modulo 4.

\section{Axion insulator NpBi}

Rare-earth monopicnitides NpX (X=As, Sb, Bi) are a very interesting class of materials that have been studied in recent years due to their unusual electronic, magnetic, and optical properties with potential applications in spintronics and high-performance computing. 
In particular, light actinide atoms such as Np exhibit 5f electronic orbitals that often hybridize with the conduction band, leading to a logarithmic decrease of the electrical resistivity with the temperature at moderate pressures \cite{PhysRevB.56.14481,LANDER19957}. 
Neptunium has several isotopes of which the most commonly found is $^{237}$Np with a half-life time of 2.14 million years. This has permitted the realization of several neptunium materials such as hydroxides \cite{Fahey1976}, oxides \cite{Richter1987, Morss1994} and halides \cite{Malm1958}. 
Very recently, it was predicted that several of these monopnictides could be axion insulators \cite{High-throughput}. 
To the best of our knowledge, and although these materials have been realized experimentally \cite{PhysRevB.56.14481,LANDER19957}, no detailed studies on topological properties have been carried out to date. In this work, we use computational methods to study the topological properties of NpBi as a case study to further illustrate the ideas discussed in the previous section.\\

NpBi is a material that crystallizes in a NaCl-type structure with a cubic Bravais lattice due to noncollinear antiferromagnetism and a N\'eel temperature of $T_N=193.5 K$ \cite{PhysRevB.56.14481}. 
The Np atoms are arranged in a tetrahedral configuration (structure known as noncollinear 3\textbf{k}) where the local magnetic moment at the ground state is around 2.6$\mu_B$ due to the Np atom \cite{PhysRevB.56.14481, BURLET1992131}. {This material
was previously predicted by high-throughput calculations \cite{High-throughput} as an axion insulator. In the present work, we go beyond these results and confirm its axion insulator behavior using additional topological indicators as it is shown in what follows.}

\subsection{Band inversion and symmetry indicators}

A sketch of the crystal and magnetic structure is shown in Figure \ref{fig1}a.
In this crystal structure, the Np atoms present an all-in/all-out (AIAO) magnetic order with a zero magnetic moment by unit-cell. The unit cell contains four Np/Bi atoms and has Pn$\overline{3}$m$’$ (\#224.113) for magnetic space group (MSG).

The symmetry group includes 48 operations generated by:
\begin{align}
 I,  \ C_3,  \ \mathbf{t}_\text{0\textonehalf \textonehalf}C_{2x}, \
\mathbf{t}_\text{\textonehalf \textonehalf0}C_{2z}  \ \  \mbox{and} \ \  \mathbb{T} C_{2 \mathbf{d}},
\end{align} 
where $C_{2x}$ and $C_{2z}$ are 2-fold rotations
around the $x$-axis and the $z$-axis respectively,  $C_{2 \mathbf{d}}$ is a 2-fold rotation around the diagonal $\mathbf{d}= (1 \bar{1} 0)$,  
$\mathbf{t}_\text{0\textonehalf \textonehalf}$ denotes translation in the direction of the vector  
$(\text{0\textonehalf \textonehalf})$, $C_3$ is a 3-fold rotation around the vector $(111)$ and $\mathbb{T}$ is the time reversal operator.
It is important to emphasize that the symmetry group does not include the time reversal symmetry.
The symmetry $ \mathbf{t}_\text{0\textonehalf \textonehalf}C_{2x}$ 
forces the Chern numbers across the planes parallel to $k_y=0$ and $k_z=0$ to be trivial, while
$\mathbf{t}_\text{\textonehalf \textonehalf0}C_{2z}$ forces the Chern numbers across 
planes parallel to $k_x=0$ to be also trivial. Therefore the Chern-Simons axion coupling of the system
is equivalent to the K-theory Kane-Mele invariant  \S \ref{section trivial Chern numbers},
and a band inversion should be expected if the material is a topological insulator \S \ref{section band inversion}. The existence of $\mathbb{Z}_2$ topological invariant for the \#224.113 MSG is consistent with the topological classification recently found by \citet{Topological-classification-and-diagnosis}

 \begin{figure}
	\includegraphics[width=8.8cm]{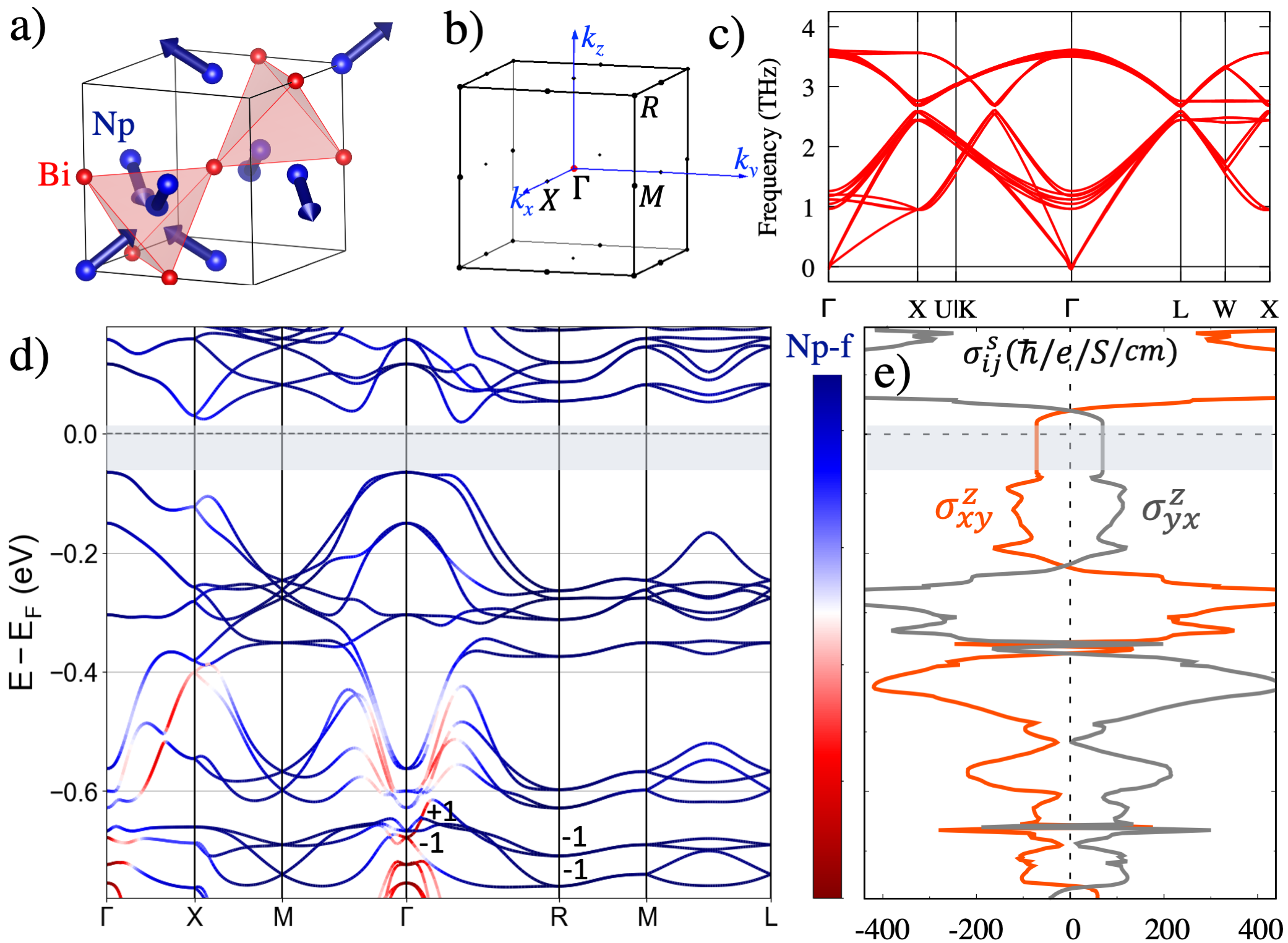}
	\caption{a) Crystal and magnetic structure of the NpBi noncollinear antiferromagnetic solid.  b) The first Brillouin zone of the cubic lattice. c) Dynamic stability from phonon structure calculations. d) Calculated bulk band structure with Np-$f$ orbital projected states on the energy bands of NpBi. A double band inversion is noted around -0.66 eV. Eigenvalues of the inversion symmetry operator change from -1 and +1 in  $\Gamma$ to +1 and +1 in $R$ point. e) Spin Hall conductivity as a function of the Fermi level. It is noted a QSHC in the bandgap of the material characterizes the topological nature of NpBi.} 
	\label{fig1}
\end{figure}

The dynamical stability of the NpBi structure is supported by the calculated vibrational dispersion curve as shown in Figure \ref{fig1}c. 
Note that the NpBi magnetic structure is similar to the pyrochlore lattice, which has been theoretical and experimentally associated with different kinds of topological states \cite{topoAFMs}. 
In particular, pyrochlore tight-binding (TB) models indicate that this magnetic structure could be a prototype to host axion insulators \cite{Surfaces_of_axion_insulators}.  
Therefore, the theoretical predictions around this TB model can be extended to the NpBi material.

On the other hand, the Np and Bi atoms in the crystal structure contribute to a strong spin-orbit coupling in the material, opening a bandgap and inducing band inversion in the valence band, as it is shown in Figure \ref{fig1}d. 
In this Figure, the Np-$f$ orbital-projected band structure is plotted along the high-symmetry lines (see Figure \ref{fig1}b). It is noted that there is a band inversion in a pair of bands around -0.66 eV. 
The band inversion is initiated in $\Gamma$ and extended to any other high symmetry point at the Brillouin zone, which induces a close surface in reciprocal space where the band inversion occurs (see Fig. \ref{Figure local Hamiltonian} and section \ref{section band inversion}). This band inversion is also linked to the change of the inversion operator eigenvalues at TRIMs as it is shown in section \ref{section inversion}.

The presence of the inversion symmetry permits to determine the K-theory Kane-Mele invariant of Eqn. \eqref{K-theory C2T} in section 
\ref{section bulk invariant} from the parity eigenvalues at high symmetry points in reciprocal space as described in Eqn. \eqref{Z4 invariant}. 
This indicator is equivalent to the $\mathbb{Z}_4$ invariant used to characterize normal, topological and axion insulators
\cite{Quantized_response_and_topology_of_magnetic_insulators, Inversion_symmetric_topological_insulators}.
To find out this indicator, we first need to notice that the information is localized only on the points
$\Gamma=(0,0,0)$ and $R=(\text{\textonehalf},\text{\textonehalf},\text{\textonehalf})$ because of symmetry considerations.

Consider the symmetry $\mathbb{T}S$ of NpBi with 
\begin{align}
S(x,y,z)=(x+\tfrac{1}{2}, z+\tfrac{1}{2},-y),
\end{align}
and note that the TRIMs $X=(\text{\textonehalf},0,0)$ and $M=(0, \text{\textonehalf},\text{\textonehalf})$
are fixed by $\mathbb{T}S$ in momentum space. Since
\begin{align}
\mathrm{t}_{110}I (\mathbb{T}S) = (\mathbb{T}S)I,
\end{align}
we have that in momentum coordinates
\begin{align}
e^{-ik_x-ik_y}I (\mathbb{T}S) = (\mathbb{T}S)I.
\end{align}
Hence, if $\Psi$ is an eigenvector of the Hamiltonian with $I \Psi = \pm \Psi$ on either $X$ or $M$, then
the eigenvector $(\mathbb{T}S)\Psi$ has opposite $I$-eigenvalues as $\Psi$. Therefore on $X$ and $M$,
the number of positive and negative $I$-eigenvalues is the same. 

On NpBi there are 120 valence bands and the distribution of positive/negative $I$-eigenvalues on $\Gamma$ and $R$ is
$62/58$ and $ 32/88$ respectively; in all the other 6 invariant points the distribution is $60/60$. Hence
the K-theory Kane-Mele invariant of the system calculated with the help of Eqn. \eqref{Z4 invariant} is 
\begin{align}
58+88 \equiv 2 \ \mathrm{mod} \ 4.
\end{align}

This suggests that the NpBi material is protected by the nontrivial topological index $\mathbb{Z}_4$=2. In  Fig. \ref{fig1}d it
could be seen that the four energy bands that localize in $\Gamma$ at around $-0.66$eV posses the non-trivial
K-theory Kane-Mele index. On $\Gamma$ there are two negative eigenvalues while on $R$ there are four. 
According to Eqn. \eqref{Z4 invariant}, either $C_{2{\bf d}}\mathbb{T}$ or inversion symmetry protects this topological 
nature of the NpBi, and the material could be considered as an axion insulator with the quantized topological 
magnetoelectric (TME) $\theta$=$\pi$.

\subsection{Gapped surface states and surface AHC}

The topological behavior of NpBi is confirmed by the quantum spin Hall conductivity (QSHC) which is maintained inside the band gap ($\sim$ 0.09eV). 
The QSHC shown in Figure \ref{fig1}e is usually associated with the presence of gapless surface states in $\mathbb{T}$ invariant systems.
However, using the Green function method calculations, no gapless surface states were found for the NpBi$(100)$,  $(1\bar{1}0)$, and $(111)$ slab structures, 
indicating that there are no symmetry operations which protect surface states in NpBi. 
In magnetic and nonmagnetic topological materials, $\mathbb{T}$ or ``effective'' $\mathbb{T}$ symmetry, protect the Dirac surface states; here it is not the case. 
The gapped surface states for the $(100)$, $(1\bar{1}0)$ and $(111)$NpBi surfaces are shown in Figure \ref{fig2}a, 
where red and blue colors indicate the weight of projections onto bottom and top surface, respectively.
These results are consistent with the gapped surface state for the $(001)$NpAs monolayer reported by \citet{Zou_2021}.

\begin{figure}
	\includegraphics[width=8.8cm]{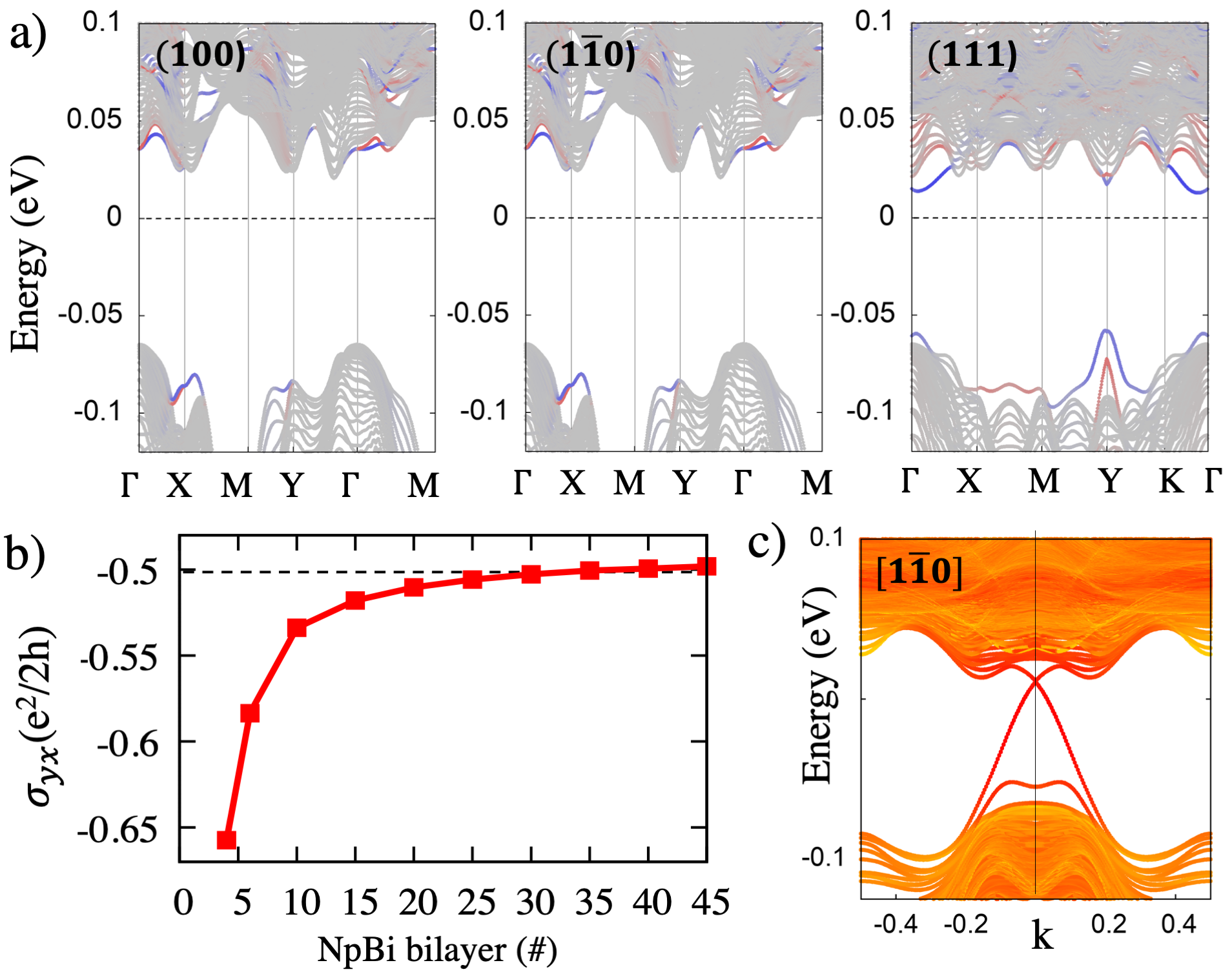}
	\caption{a) Calculated band structure of the $(100)$, $(1\bar{1}0)$, and $(111)$NpBi surface. Red and blue colors represent the projections onto the bottom and top surfaces respectively. Bulk projected states are indicated in gray color. b) Surface anomalous Hall conductivity (per slab thickness) as a function of the number of Np-Bi bilayers. A half-quantized anomalous Hall conductivity at the $(111)$ surface is noted, which implies an axion insulator state for NpBi. c) Gapless chiral hinge states calculated for the NpBi nanowire grown in the $(1\bar{1}0)$ direction.} 
	\label{fig2}
\end{figure}

In order to confirm the axion insulator phase of the NpBi material, we have calculated the slab Berry curvature $\Omega_z(k)$ for the $(100)$, $(1\bar{1}0)$ and $(111)$ surfaces using the {\tt WannierTools} code \cite{wanniertools}.  
Then, the integral of the curvature $\Omega_z(k)$ over the surface Brillouin zone was used to calculate the surface anomalous Hall conductivity (sAHC).  
We found trivial sAHC for the $(100)$ and $(1\bar{1}0)$ surfaces, due to the cancellation of the  integral of the curvature $\Omega_z(k)$ by surface symmetries.
However, it is found that the Berry curvature perpendicular to the $(111)$ surface  is nonzero, not compensated, and protected by the $C_3$ surface symmetry.

The (111)NpBi surface consists of alternating Bi and Np hexagonal monolayers with a 2D hexagonal Brillouin zone.  
In this surface we found a saturation value of $e^2/2h$ for the sAHC (per slab thickness).  
A saturation behavior of the half-quantized sAHC (hq-sAHC) as a function of the number of NpBi bilayers is plotted in Figure \ref{fig2}b.   
Slabs with thickness more than 20  bilayers are enough to display the half-quantized sAHC for NpBi$(111)$ surface in the absence of an external magnetic field. 
We also found a sign inversion of the hq-sAHC when the $(\bar{1}11)$ surface is used. 
This signal indicates that the NpBi is actually an axion insulator that exhibits a quantized bulk TME manifested by means the hq-sAHC.

A similar prediction for the gapped surface states and hq-sAHC has been carried out for the $(111)$ surface of the pyrochlore lattices
 \cite{Surfaces_of_axion_insulators}, which is similar to the NpBi structure. 
In addition, it is important to mention that the magnetic Np atoms in the layers of the $(111)$NpBi slab form the known 3Q noncoplanar antiferromagnetic structure, which is a noncollinear spin arrangement on a two-dimensional lattice. 
This configuration has been observed in Mn/Cu$(111)$ surfaces \cite{MnCu111} and leads to anomalous Hall and topological Hall effects \cite{3QAFM}. 

A comparison between the potential axion insulator materials is shown in Table \ref{axioninsulators}.
The axion insulator behavior for the protected $C_{2{\bf d}}\mathbb{T}$ symmetry is different from the $t_{\text{\textonehalf}} \mathbb{T}$ ($\mathbb{T}$ composed partial translation of the unit cell) and $C_{2y} \mathbb{T}$ systems. 
For the $t_{\text{\textonehalf}} \mathbb{T}$ cases, gapless surface states are observed in TRIMs for MnBi$_2$Te$_4$ surfaces that contain the partial translation vector $t_{\text{\textonehalf}}$ \cite{EuIn2As2-Riberolles}.  
For the $C_{2y} \mathbb{T}$ symmetry, EuIn$_2$As$_2$, EuSn$_2$As$_2$ and EuSn$_2$P$_2$ could show protected gapless surface states at the $(010)$ surfaces; this matter is a subject of intense research \cite{EuIn2As2-prr, EuIn2As2-prl, EuSn2As2, EuSn2P2-Gui, EuSn2P2-pnas}.
However, we found gapped surfaces states for all NpBi surfaces, thus making NpBi a topological axion insulator without surface states.

\begin{table*}
	\begin{tabular}{c|cccccccccc} 
		\hline \hline
		\multirow{2}{*}{Material} & \multirow{2}{*}{Symmetry} & \multirow{2}{*}{Magnetization} & Bandgap & Band & Topological & Surface & Chiral hinge  & \multirow{2}{*}{QSHC} & \multirow{2}{*}{hq-sAHC} & \multirow{2}{*}{Reference} \\ 
		 &  &  &  (eV) & inversion & invariant & states & states  &  &  & \\ 
		\hline
		MnBi$_2$Te$_4$   & $t_{1/2} \mathbb{T}$ & CAFM & $\sim$0.16 & $\pmb{\checkmark}$ & $\pmb{\checkmark}$ & $\pmb{-}$ & $\pmb{-}$ & $\pmb{\checkmark}$ & $\pmb{-}$  & \cite{MnBi2Te4-Gu,MnBi2Te4-family,MnBi2Te4-jing} \\
		MnBi$_6$Te$_{10}$   & $t_{1/2} \mathbb{T}$ & CAFM & $\sim$0.20 & $\pmb{\checkmark}$ & $\pmb{\checkmark}$ & $\pmb{X}$  & $\pmb{-}$ & $\pmb{\checkmark}$ & $\pmb{-}$ &  \cite{MnBi6Te10, MnBi6Te10-tian} \\
		MnBi$_8$Te$_{13}$   & $t_{1/2} \mathbb{T}$ &  FM  & $\sim$0.17 & $\pmb{\checkmark}$ & $\pmb{\checkmark}$ & $\pmb{X}$  & $\pmb{-}$ & $\pmb{\checkmark}$ & $\pmb{-}$ & \cite{MnBi8Te13,MnBi8Te13-zhong}  \\		
		EuIn$_2$As$_2$& $C_{2y} \mathbb{T}$ & CAFM  & $\sim$0.10 & $\pmb{\checkmark}$ & $\pmb{\checkmark}$ & $\pmb{-}$ & $\pmb{\checkmark}$ & $\pmb{\checkmark}$ & $\pmb{-}$ &  \cite{EuIn2As2-prr,EuIn2As2-prl} \\
		EuSn$_2$As$_2$ & $C_{2y} \mathbb{T}$ & CAFM  & $\sim$0.13  & $\pmb{\checkmark}$ & $\pmb{\checkmark}$ & $\pmb{-}$ & $\pmb{\checkmark}$ & $\pmb{\checkmark}$ & $\pmb{-}$ &\cite{EuSn2As2,EuSn2As2-lv}   \\
		EuSn$_2$P$_2$ & $t_{1/2} \mathbb{T}$ & CAFM  &   $\pmb{-}$ & $\pmb{\checkmark}$ & $\pmb{\checkmark}$ & $\pmb{-}$ & $\pmb{\checkmark}$ & $\pmb{\checkmark}$ & $\pmb{-}$  & \cite{EuSn2P2-Gui, EuSn2P2-pnas} \\
		NpBi  & $C_{2{\bf d}} \mathbb{T}$ & NCAFM  & $\sim$0.09 & $\pmb{\checkmark}$ & $\pmb{\checkmark}$ & $\pmb{X}$ & $\pmb{\checkmark}$ & $\pmb{\checkmark}$ & $\pmb{\checkmark}$  & \\ 
		\hline \hline
	\end{tabular} \label{axioninsulators}
	\caption{Axion insulators materials with their symmetry protected topological phase. $t_{1/2}$ is a translation operation of half unit cell along the $z$ axis and $C_{2y}\mathbb{T}$ corresponds to combined $C_{2y}$ two-fold rotation around $y$-axis with time reversal symmetry. Ferromagnetic (FM), collinear (CAFM) or noncollinear (NCAFM) antiferromagnetic stable structures are indicated. Bandgap energy (in eV) and band inversion phenomena in the valence band. Topological invariant index calculated from generalized Fu-Kane-Mele inversion symmetry eigenvalue formula. Surface states and hinge states. Quantum spin Hall conductivity (QSHC) in the bulk material and half-quantized surface anomalous Hall conductivity (hq-sAHC). Existence of surface states in MnBi$_2$Te$_4$, EuIn$_2$As$_2$, EuSn$_2$As$_2$ and EuSn$_2$P$_2$ is still a matter of current research. MnBi$_2$Te$_4$ is a layered material, which can be classified as axion insulator according to the number of septuple-layers.} 
	\label{axioninsulators}
\end{table*}

\subsection{Chiral hinge states}

The observation of QSHC in the bandgap and the absence of gapless surface states could indicate that the NpBi can manifest a bulk-surface-hinge instead of a bulk-surface correspondence. 
This phenomenon has been predicted for the axion insulators and it has been also observed in the called higher order topological insulators (HOTIs) \cite{HOTI}. 
We have found that the NpBi could host gapless topological states along the hinges as it was confirmed by the band structure of a nanowire growth in the $[1\bar{1}0]$ direction shown in Figure \ref{fig2}c.  
The nanowire supercell consists of a 11$\times$11$\times$1 unit cells periodic along the $[1\bar{1}0]$ direction surrounded by vacuum space (trivial insulator) in the perpendicular directions.

From Figure \ref{fig2}c, it is demonstrated the existence of chiral hinge states inside the bandgap energy, which are extended along the NpBi
 $[1\bar{1}0]$ direction.  
The existence of one-dimensional hinge states in the rotational invariant directions $[1\bar{1}0]$ is preserved by the $C_{2{\bf d}}\mathbb{T}$ symmetry of the magnetic structure of NpBi. 
The twelve equivalent $\langle 1\bar{1}0 \rangle$ directions of the hinge states correspond to the intersection of the family of eight $\{111\}$ gapped surfaces, where the hq-sAHC is observed.

In three dimensions, the chiral hinge modes stabilized by the $C_{2{\bf d}}\mathbb{T}$ and $C_3$ symmetries exhibit an octahedral crystal shape in real space as it is shown the Figure \ref{fig3}. 
In this figure we have plotted with red and blue the gapped surfaces according to the sign of the calculated hq-sAHC.
Black arrows indicate the circulation of the hinge state, where clockwise rotation corresponds to the blue faces with positive hq-sAHC. 
A similar octahedral crystallite configuration for chiral hinge states was also reported for the AIAO spin configuration model by \citet{Surfaces_of_axion_insulators}. 
Finally, it is important to remark that the gapless chiral hinge states of the NpBi material can carry out dissipationless charge/spin transport,  which furthermore confirms the presence of QSHC in the bulk and hq-sAHC in the surfaces.
 
\begin{figure}
	\includegraphics[width=4.8cm]{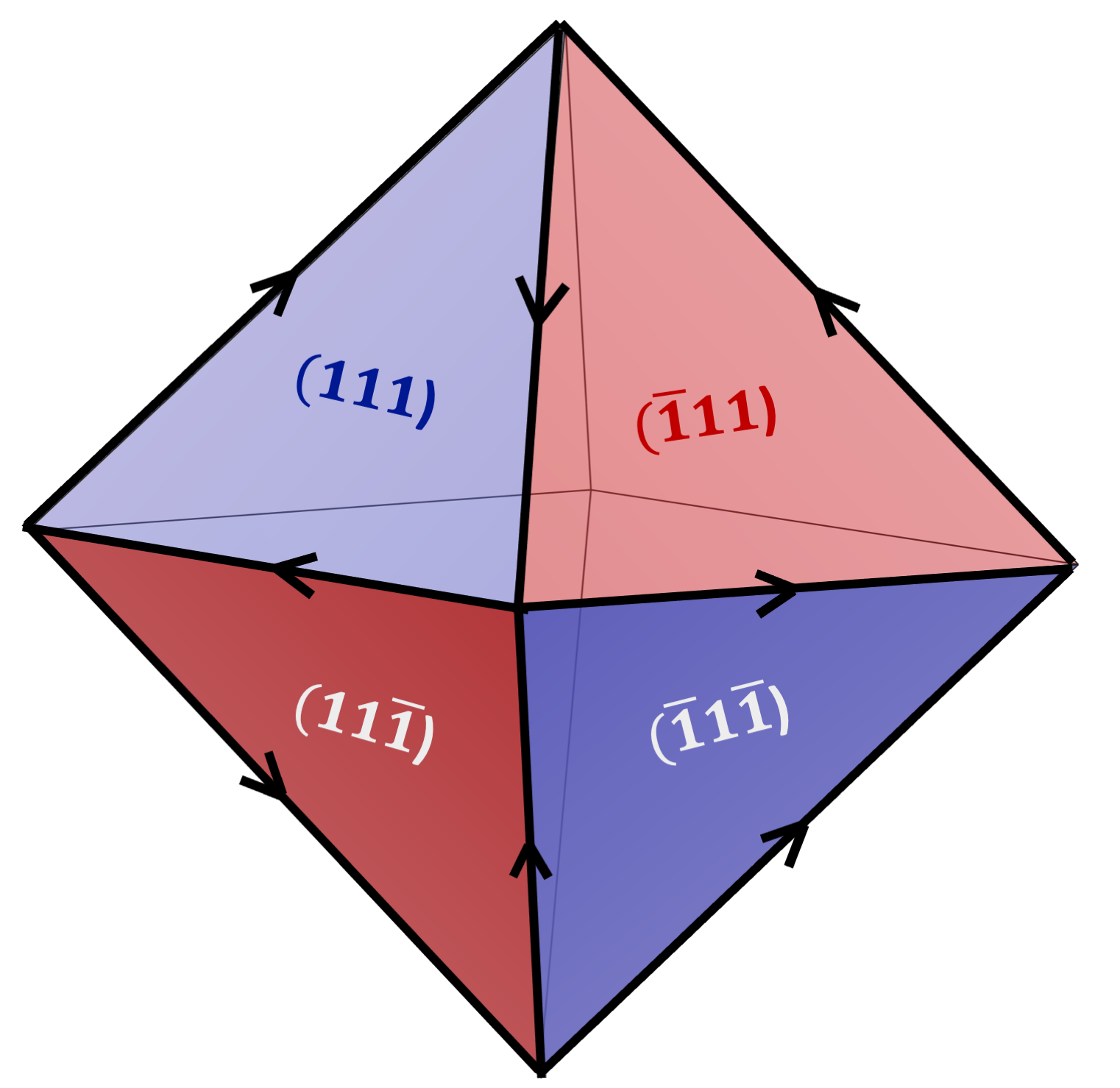}
	\caption{Gapless hinge states for the equivalent  $\langle 1\bar{1}0 \rangle$ directions of the NpBi material. Chiral hinge states generate an octahedral macroscopic crystal with eight faces which correspond to the equivalent $\{111\}$ surfaces. Blue and red surfaces indicate positive and negative hq-sAHC. Black arrows show the circulation of the chiral hinge state, where clockwise rotation corresponds to positive hq-sAHC.} 
	\label{fig3}
\end{figure}

\subsection{Computational method}

First-principles calculations were performed within the density functional theory (DFT) formalism, as implemented in the Vienna \textit{ab-initio}  simulation package (VASP) \cite{vasp}. 
Exchange-correlation effects were included using the generalized gradient approximation (GGA) with the Perdew-Burke-Ernzerhof (PBE) parametrization \cite{pbe}.
Electron-ion interactions were treated with the projected augmented wave (PAW) method \cite{PAW}, and considering Np-$f$ electrons as valence states, without the Hubbard correction.
Spin–orbit coupling was included in the DFT calculations.
The cutoff parameter for the electronic wave functions was set to be 520 eV. 
A $\Gamma$-centered grid of 10$\times$10$\times$10 {\bf k}-point was used to sample the Brillouin zone (BZ). 
Projected band structures were plotted using the {\tt pyprocar} code \cite{pyprocar}.
The phonon spectrum was obtained by the {\tt phonopy} code within the density function perturbation theory \cite{phonopy} using a 2$\times$2$\times$2 \textbf{q}-point grid.  
The Fu-Kane-Mele invariant was calculated from the inversion symmetry eigenvalues at the high-symmetry points of the BZ using the {\tt irrep} package \cite{irrep}.

By projecting the Bloch states into Wannier functions (WF) (Np-$f$ and Bi-$p$), we built a tight-binding model based on maximally localized Wannier functions (MLWF) \cite{wannier90}. 
The surface states and surface Berry curvature were computed using a 121$\times$121 {\bf k}-mesh into the iterative Green’s function approach as implemented in the  {\tt WannierTools} package \cite{wanniertools}.  
For slab calculations, an sample with 31-unit-cell thick in the non-periodic direction was used.
Finally, we calculated the intrinsic spin Hall conductivity (SHC) by integrating the spin Berry curvature in a 320$\times$320$\times$320 {\bf k}-grid using the  {\tt linear-response} code \cite{SHEcode}.

\section{Acknowledgments}
R.G-H. gratefully acknowledges the computing time granted on the supercomputer Mogon at Johannes Gutenberg University Mainz \footnote{hpc.uni-mainz.de}. 
C.P. thanks the support of the Science and Technology Facilities Council of the UK for computing time on SCARF.  
B.U acknowledges the support of CONACYT through Proj. No. CB-2017-2018-A1-S-30345-F-3125 and 
 of the Max Planck Institute for Mathematics at Bonn, Germany. R.G-H. and B.U. gratefully acknowledge the continuous support of the Alexander Von Humboldt Foundation, Germany. R.G-H., C.P. and B.U. acknowledge the support from the ICTP through the Associates Programme.


\bibliographystyle{apsrev}
\bibliography{bibliography}

\end{document}